\documentclass[12pt,draftclsnofoot,onecolumn]{IEEEtran}

\usepackage[pdftex]{graphicx}

\usepackage[numbers]{natbib}
\let\chapter\section
\usepackage[printonlyused]{acronym}
\usepackage[algoruled]{algorithm2e}
\usepackage{amsmath}
\usepackage{amssymb}
\usepackage{amsthm}
\usepackage{balance}
\usepackage{color}
\usepackage{float}
\usepackage{etoolbox}
\usepackage{framed}
\usepackage{cleveref} 
\usepackage{listings}
\usepackage{MnSymbol}
\usepackage{multirow}

\usepackage{setspace}
\usepackage{subfig}
\usepackage{verbatim}
\usepackage{xcolor}

\definecolor{myRed}{rgb}{1,0,0}
\definecolor{myGreen}{rgb}{0,1,0}
\definecolor{myBlue}{rgb}{0,0,1}
\crefformat{chapter}{#2Chapter~#1#3}
\crefformat{section}{#2Section~#1#3}
\crefformat{subsection}{#2Section~#1#3}
\crefformat{appendix}{#2Appendix~#1#3}
\crefformat{figure}{#2Figure~#1#3}
\crefformat{table}{#2Table~#1#3}
\crefformat{lemma}{#2Theorem~#1#3}
\crefformat{algorithm}{#2Algorithm~#1#3}
\crefformat{equation}{#2(#1)#3}
\crefmultiformat{equation}{#2(#1)#3}{ and~#2(#1)#3}{, ~#2(#1)#3}{, and~#2(#1)#3}
\crefmultiformat{figure}{#2Figures~#1#3}{ and~#2#1#3}{, ~#2#1#3}{, and~#2#1#3}
\crefmultiformat{table}{#2Tables~(#1)#3}{ and~#2(#1)#3}{, ~#2(#1)#3}{, and~#2(#1)#3}
\crefmultiformat{section}{#2Sections~#1#3}{ and~#2#1#3}{, ~#2#1#3}{, and~#2#1#3}
\crefrangeformat{equation}{#3(#1)#4--#5(#2)#6}
\crefrangeformat{figure}{#3Figures~#1#4--#5#2#6}
\crefrangeformat{section}{#3Sections~#1#4--#5#2#6}
\crefrangeformat{algorithm}{#3(#1)#4--#5(#2)#6}
\crefname{algocf}{Algorithm}{Algorithms}
\Crefname{algocf}{Algorithm}{Algorithms}
\crefname{appsec}{Appendix}{Appendices}

\newcommand{\Exp}{\mathbb{E}}
\newcommand{\fa}{\mathbf{a}}
\newcommand{\fb}{\mathbf{b}}
\newcommand{\fk}{\mathbf{k}}
\newcommand{\fT}{\boldsymbol{\Theta}}
\newcommand{\ftheta}{\boldsymbol{\theta}}

\newcommand{\mK}{\mathcal{K}}
\newcommand{\mT}{\mathcal{T}}
\newcommand{\mS}{\mathcal{S}}
\newcommand{\mI}{\mathcal{I}}
\newcommand{\mA}{\mathcal{A}}
\newcommand{\mN}{\mathcal{N}}
\newcommand{\NI}{N_{\rm I}}
\newcommand{\Rin}{R_{\rm in}}
\newcommand{\Rout}{R_{\rm out}}

\newcommand{\customwidthsmall}{.6\columnwidth}

\newcommand{\mydoubleW}{.45\columnwidth}

\DeclareSymbolFont{extraup}{U}{zavm}{m}{n}
\DeclareMathSymbol{\vardiamond}{\mathalpha}{extraup}{87}

\newtheorem{theorem}{Theorem}[section]

\lstnewenvironment{mat}{\lstset{language=mathematica,mathescape,columns=flexible}}{}
\colorlet{shadecolor}{blue!5}

\acrodef{3GPP}{3rd Generation Partnership Project}
\acrodef{BER}{Bit Error Ratio}
\acrodef{BS}{Base Station}
\acrodef{CCI}{Co-Channel Interference}
\acrodef{CDF}{Cumulative Distribution Function}
\acrodef{CF}{Characteristic Function}
\acrodef{CIR}{Carrier to Interference Ratio}
\acrodef{CoMP}{Coordinated Multi-Point}
\acrodef{ECDF}{Empirical Cumulative Distribution Function}
\acrodef{eICIC}{Enhanced Inter-Cell Interference Coordination}
\acrodef{GIG}{Generalized Integer Gamma}
\acrodef{ICIC}{Intercell Interference Coordination}
\acrodef{LTE}{Long Term Evolution}
\acrodef{MGF}{Moment Generating Function}
\acrodef{MIMO}{Multiple Input Multiple Output}
\acrodef{MISO}{Multiple Input Single Output}
\acrodef{MRC}{Maximum Ratio Combining}
\acrodef{MRT}{Maximum Ratio Transmission}
\acrodef{MU}{Multi-User}
\acrodef{SVD}{Singular Value Decomposition}
\acrodef{PAPR}{Peak-to-Average Power Ratio}
\acrodef{PPP}{Poisson Point Process}
\acrodefplural{PPP}[PPPs]{Poisson Point Processes}
\acrodef{PDF}{Probability Density Function}
\acrodef{PP}{Point Process}
\acrodef{RV}{Random Variable}
\acrodef{SIR}{Signal-to-Interference Ratio}
\acrodef{SISO}{Single Input Single Output}
\acrodef{ZF}{Zero Forcing}

\begin{document}

\title{ A Circular Interference Model for \\ Asymmetric Aggregate Interference}
\author{
\IEEEauthorblockN{Martin Taranetz, Markus Rupp}\\
\IEEEauthorblockA{Vienna University of Technology, Institute of Telecommunications \\
Gusshausstrasse 25/E389, A-1040 Vienna, Austria\\
Email:  \{mtaranet,mrupp\}@nt.tuwien.ac.at
}}
\maketitle

\begin{abstract}

Scaling up the number of base stations per unit area is one of the major trends in mobile cellular systems of the fourth (4G)- and fifth generation (5G), making it increasingly difficult to characterize aggregate interference statistics with system models of low complexity. This paper proposes a new circular interference model that aggregates given interferer deployments to power profiles along circles. The model accurately preserves the original interference statistics while considerably reducing the amount of relevant interferers. In comparison to common approaches from stochastic geometry, it enables to characterize cell-center- and cell-edge users, and preserves effects that are otherwise concealed by spatial averaging. To enhance the analysis of given power profiles and to validate the accuracy of the circular model, a new finite sum representation for the sum of Gamma random variables with integer-valued shape parameter is introduced. The approach allows to decompose the distribution of the aggregate interference into the contributions of the individual interferers. Such knowledge is particularly expedient for the application of base station coordination- and cooperation schemes. Moreover, the proposed approach enables to accurately predict the corresponding signal-to-interference-ratio- and rate statistics.

\end{abstract}



\acresetall

\section{Introduction and Contributions}

In mobile cellular systems of the fourth (4G)- and fifth generation (5G), the number of \acp{BS} per unit area is expected to grow substantially~\citep{6736747}. One of the main performance limiting factors in such dense networks is \emph{aggregate interference}. Hence, its accurate statistical characterization becomes imperative for network design and analysis. Although abstraction models such as the Wyner model and the hexagonal grid have been reported two \citep{340450}- or even five decades ago \citep{BellLabsIdeaFactory}, mathematically tractable interference statistics are still the exception rather than the rule. Moreover, the emerging network topologies fundamentally challenge various time-honored aspects of traditional network modeling~\citep{ghosh2012heterogeneous}.


In current literature, \ac{BS} deployment models mainly follow the trend away from being fully deterministic towards complete spatial randomness \citep{6171992,6779696}. 
However, even the new approaches only yield known expression for the \ac{PDF} of the aggregate interference, if particular combinations of spatial node distributions, path loss models and user locations are given~\citep{elsawy2013stochastic}.
For example, a finite, typically small number of interferers together with certain fading distributions, such as Rayleigh, lognormal or Gamma allows to exploit literature on the sum of \acp{RV}~ \citep{966578,1556824,6292935,C:Sagias_C2_2006,1096243,1673666,1583918,747812,729390,4524273,TorrieriV12,moschopoulos,1962,3717,693785,Coelho199886,abu1994outage,beaulieu1995estimating,hu2005accurate,mehta2007approximating}. 

Otherwise, tractable interference statistics have mainly been reported in the field of stochastic geometry. This powerful mathematical framework recently gained momentum as the only available tool that provides 
a rigorous approach to modeling, analysis and design of networks with a substantial amount of nodes per unit area~\citep{Baccelli96stochasticgeometry,895048,StochasticGeomAndArchi,Haenggi:2009,4802198,BaccelliVolI,BaccelliVolII,5621983,6747823,guo2013spatial,6516171,5226957,haenggi2012stochastic,NET-032,BaiBlockage}. However, when closed-form expressions are desired, it imposes its own particular limitations, typically including spatial stationarity and isotropy of the scenario~\citep{elsawy2013stochastic,Haenggi:2009,5226957}. Hence, the potential to consider an asymmetric interference impact is very limited and notions such as \emph{cell-center} and \emph{cell-edge} are, in general, not accessible. 
The contributions of this paper outline as follows:

\begin{itemize}

\item A new \emph{circular interference model} is introduced. The key idea is to map arbitrary out-of-cell interferer deployments onto circles of uniformly spaced nodes such that the original aggregate interference statistics can accurately be reproduced. The model greatly reduces complexity as the number of participating interferers is significantly reduced.


\item A \emph{mapping scheme} that specifies a procedure for determining the power profiles of arbitrary interferer deployments is proposed. Its performance is evaluated by means of Kolmogorov-Smirnov statistics. The test scenarios are modeled by \acp{PPP} so as to confront the regular circular structure with complete spatial randomness. It is shown that the individual spatial realizations exhibit largely diverging power profiles.

\item A new finite sum representation for the \ac{PDF} of the \emph{sum of Gamma \acp{RV} with integer-valued shape parameter} is introduced to further enhance and validate interference analysis with the circular model. Its restriction to integer-valued shape parameters is driven by relevant use cases for wireless communication engineering and the availability of \emph{exact} solutions. 
The key strength of the proposed approach lies in the ability to decompose the interference distribution into the contributions of the individual interferers.

\item Statistics of aggregate interference with \emph{asymmetric interference impact} are investigated. The asymmetry is induced by eccentrically placing a user in a generic, isotropic scenario. This setup is achieved by applying the introduced circular model with uniform power profiles. On top of that, the model enables to employ the proposed finite sum representation. It is shown that the partition of the interference distribution is particularly useful to identify candidate \acp{BS} for user-centric \ac{BS} collaboration schemes. Moreover, the framework allows to predict the corresponding \ac{SIR}- and rate statistics. 



\end{itemize}
This paper is organized as follows. \cref{Sec:SGN_CircleModel,Sec:SGN_SumOfGammaTheorem} introduce the circular interference model and the new finite sum representation for the sum of Gamma \acp{RV} with integer-valued shape parameter, respectively. 
\cref{Sec:SGN_SamplingArbitraryNetworks} presents a mapping scheme and validates the applicability of the circular model.  \cref{Sec:SGN_GenericScenario} investigates aggregate interference statistics and the performance of \ac{BS} collaboration schemes at eccentric user locations. \cref{Sec:SGN_Summary} concludes the work. The main focus of this paper is on downlink transmission in cellular networks. A comparable framework for the uplink is found in \citep{6378490}.

\section{Circular Interference Model}
\label{Sec:SGN_CircleModel}

\begin{figure}
	\centering
	\includegraphics[width=.6\columnwidth]{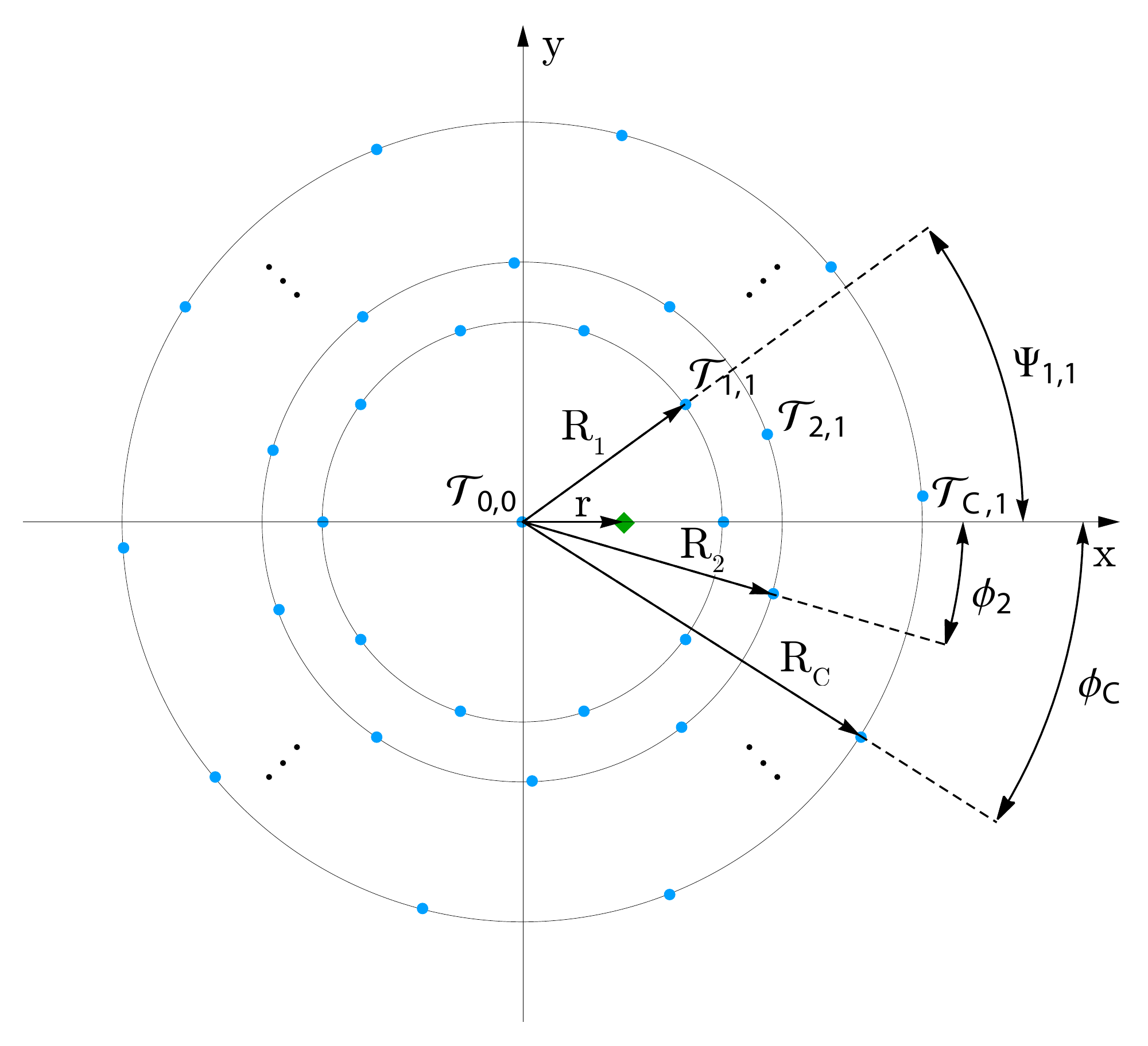}
	\caption{Circular interference model with $C$ circles of radius $R_c$ and phase $\phi_c$, $c\in\{1,\ldots,C\}$, and user at $(r,0)$. $\mT_{c,n}$ denotes the nodes of the model.}
	\label{Fig:JRingScenario}
\end{figure}

Consider the serving \ac{BS} to be located at the origin.
The proposed circular interference model is composed of $C$ concentric circles of interferers, as shown in \cref{Fig:JRingScenario}. On circle $c\in\{1,\ldots,C\}$ of radius $R_c$, $N_c$ interfering nodes are spread out equidistantly. The interferer locations are expressed in terms of polar coordinates as $(R_c,\Psi_{c,n})$, where $\Psi_{c,n} = 2\pi n/N_c-\phi_c$, with $n\in\{1,\ldots,N_c\} $  and $\phi_c \in[0,2\pi)$. Each node is unambiguously assigned to a tuple $(c,n)$ and labeled as $\mT_{c,n}$. The central \ac{BS} is denoted as  $\mT_{0,0}$. Some of the interferers on the circles may also become serving nodes when \ac{BS} collaboration schemes are applied, as will be shown later in \cref{Sec:SGN_TransmitterCollabSchemes}.

\begin{figure}
	\centering
	\includegraphics[width=.75\textwidth]{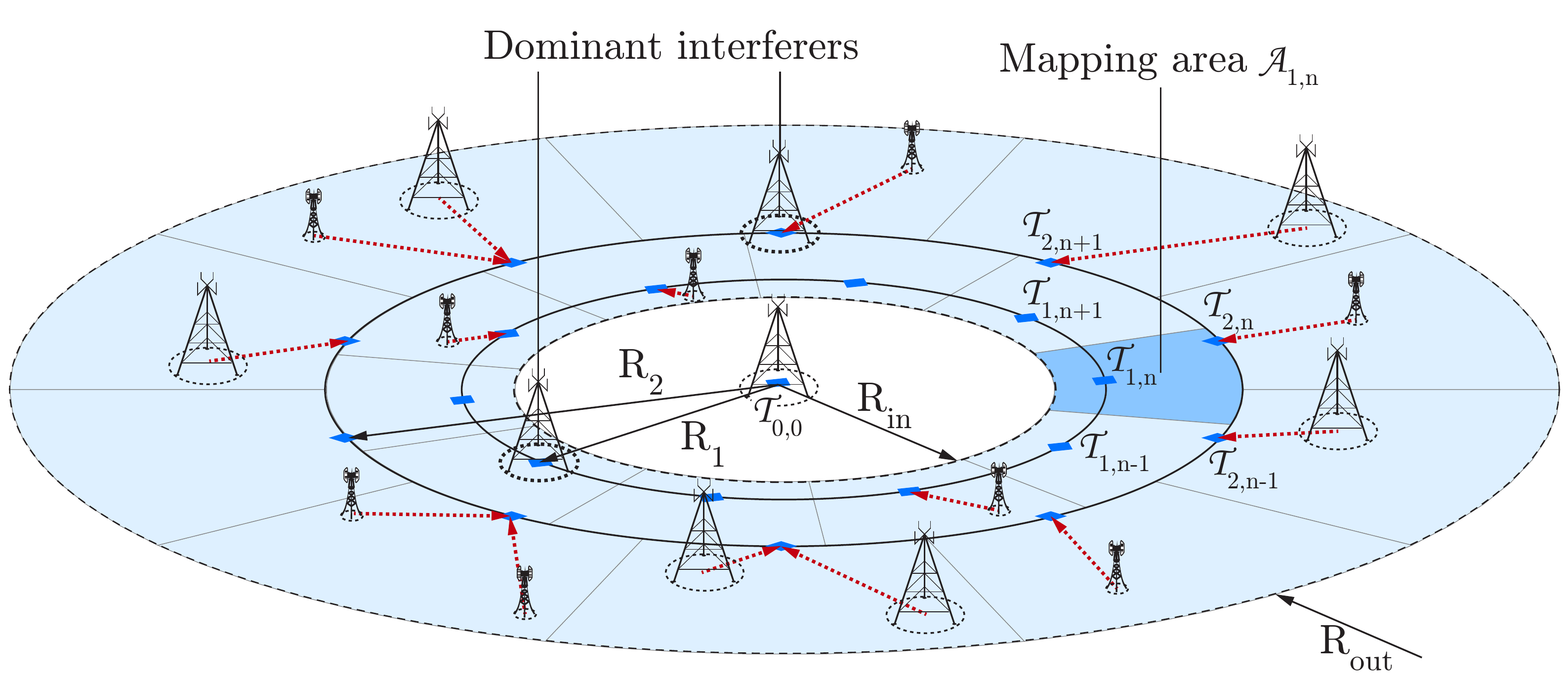}
	\caption{Circular interference model with two circles, i.e., $C=2$. Characteristics of an arbitrary heterogeneous interferer deployment are condensed to circles of equidistantly spaced nodes $\mT_{c,n}$ such that the original interference statistics can accurately be reproduced. A mapping scheme is presented in \cref{Sec:SGN_SamplingArbitraryNetworks}. The original \acp{BS} are distributed within an annulus of inner radius $\Rin$ and outer radius $\Rout$.}
	\label{Fig:CircularModelExample}
\end{figure}

The interferers on the circles do not necessarily represent real physical sources. As illustrated in \cref{Fig:CircularModelExample}, they rather correspond to the $N_c$ mapping points of an angle-dependent \emph{power profile} $p_c[n]$, with $\sum_{n=1}^{N_c} p_c[n] = 1$.  Exemplary profiles of a single circle are shown in \cref{Fig:PowerProfiles}. Intuitively, $p_c[n]$ condenses the interferer characteristics of an annulus with inner radius $\Rin$ and possibly infinite outer radius $\Rout$ such that the circular model equivalently reproduces the original \ac{BS} deployment in terms of interference statistics. This technique enables to represent substantially large networks by a \emph{finite- and well-defined} constellation of nodes. By reducing the number of relevant interferers, it greatly reduces complexity and thus allows to apply finite sum-representations as those introduced in \cref{Sec:SGN_SumOfGammaTheorem}. 

\begin{table}
	\centering 
	\caption{Parameters of the circular interference model.}
	\begin{tabular}{ r | l  }
	\hline
	\textbf{Symbol} & \textbf{Annotation} \\
	\hline
	$\Rin$ & Inner radius of mapping region, $\Rin \geq 0$ \\
	$\Rout$ & Outer radius of mapping region, $\Rout > \Rin $ \\
	\hline
	$C$ & Number of interferer circles, $C\in\mathbb{N}^+$\\
	$R_c$ & Radius of circle $c$, $c \in \{1,\ldots,C\}$, $R_c > 0 $\\
	$\phi_c$ & Phase of circle $c$,  $c\in\{1,\ldots,C\}$ $\phi_c\in[-\frac{\pi}{N_c}, \frac{\pi}{N_c}]$\\
	$N_c$ & Number of mapping points, $c\in\{1,\ldots,C\}$, $N_c\in\mathbb{N}^+$\\
	\hline
	$P_c$ & Total transmit power of circle $c$, $c\in\{1,\ldots,C\}$, $P_c >0$\\
	$p_c[n]$ &Power profile of circle $c$,  $c\in\{1,\ldots,C\}$, $n\in\{1,\ldots,N_c\}$, $p_c[n]\in[0,1]$\\
	\hline
	\end{tabular}	
	\label{Tab:SGN_Notation}
\end{table}

\cref{Tab:SGN_Notation} summarizes the parameters of the model. Typically, the size of the mapping region, as specified by $\Rin$ and $\Rout$, is predetermined by the scenario. The freely selectable variables are the amount of circles $C$ and, for each circle, the phase $\phi_c$, the radius $R_c$ and the number of mapping points $N_c$, respectively. \cref{Sec:SGN_SamplingArbitraryNetworks} presents systematic experiments to provide a reference for the parameter setting and proposes a mapping scheme to determine power profiles $p_c[n]$ and transmit powers $P_c$, respectively.

\newcommand{\myDoubleW}{.4\columnwidth}

\begin{figure}
	\centering
	\subfloat[10 interferers]{
       	\includegraphics[width = \myDoubleW]{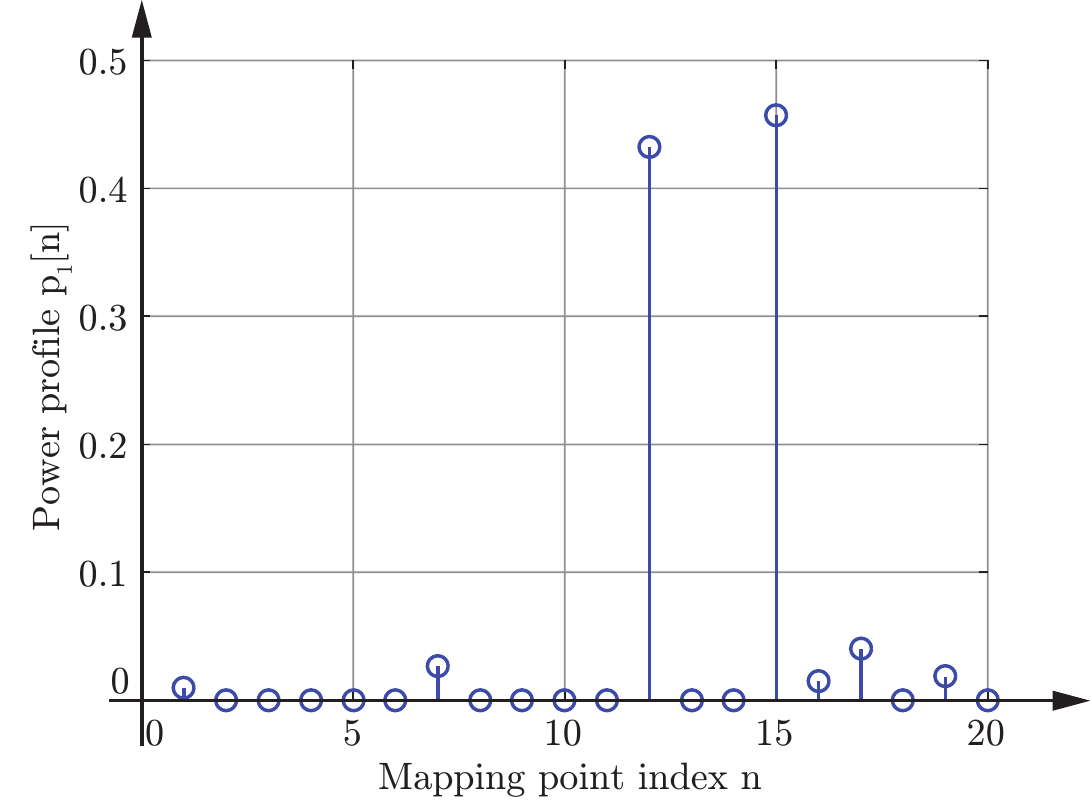}
		\label{subfig:SGN_PP_10}
	}
	\subfloat[100 interferers]{
       	\includegraphics[width = \myDoubleW]{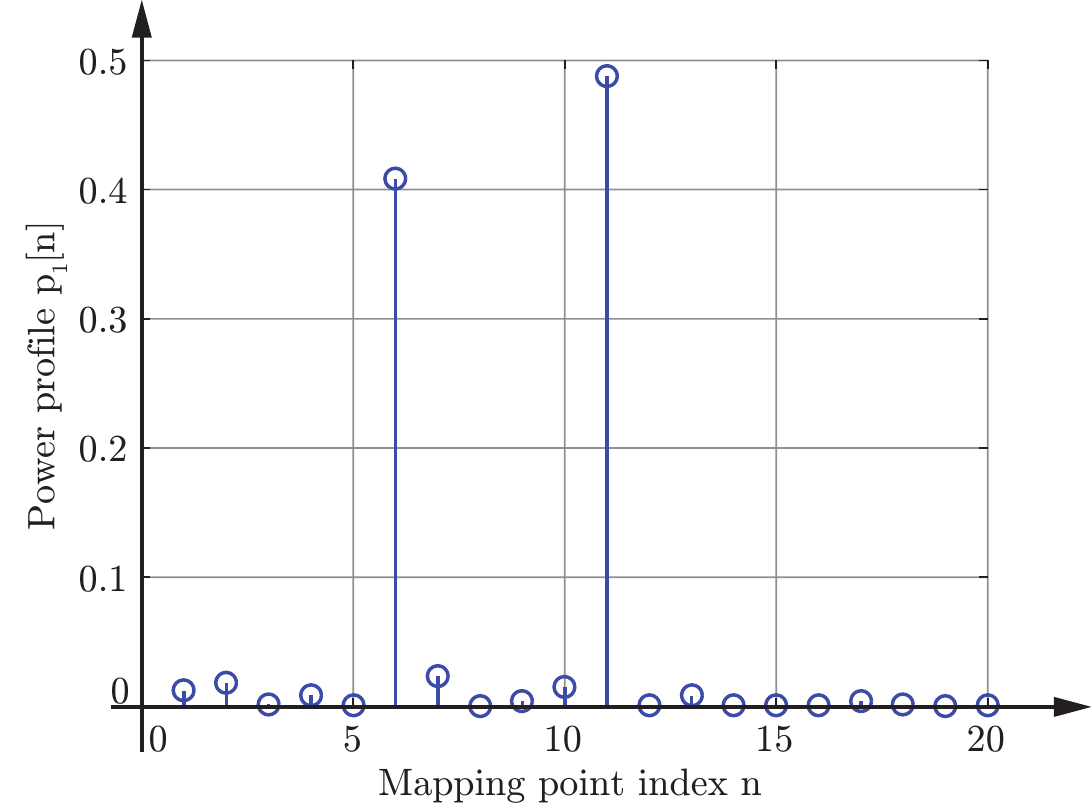}
		\label{subfig:SGN_PP_100}
	}\\
	\subfloat[1\,000 interferers]{
       	\includegraphics[width = \myDoubleW]{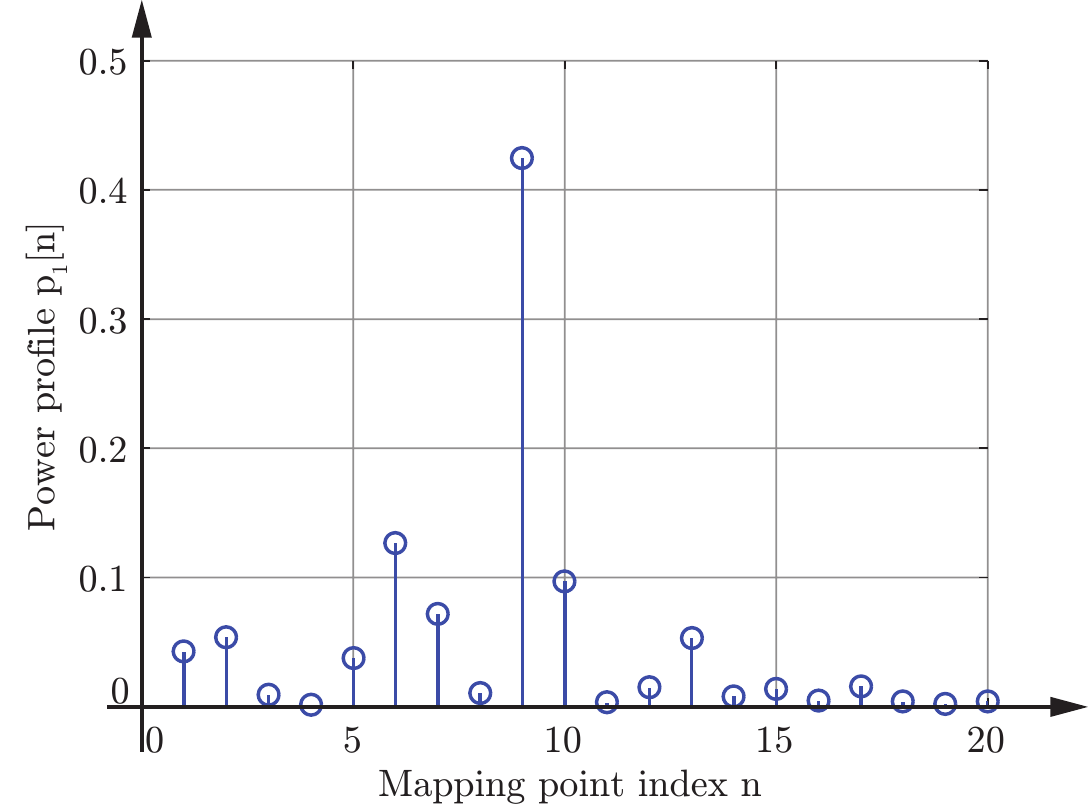}
		\label{subfig:SGN_PP_1000}
	}
	\subfloat[Hexagonal grid]{
       	\includegraphics[width = \myDoubleW]{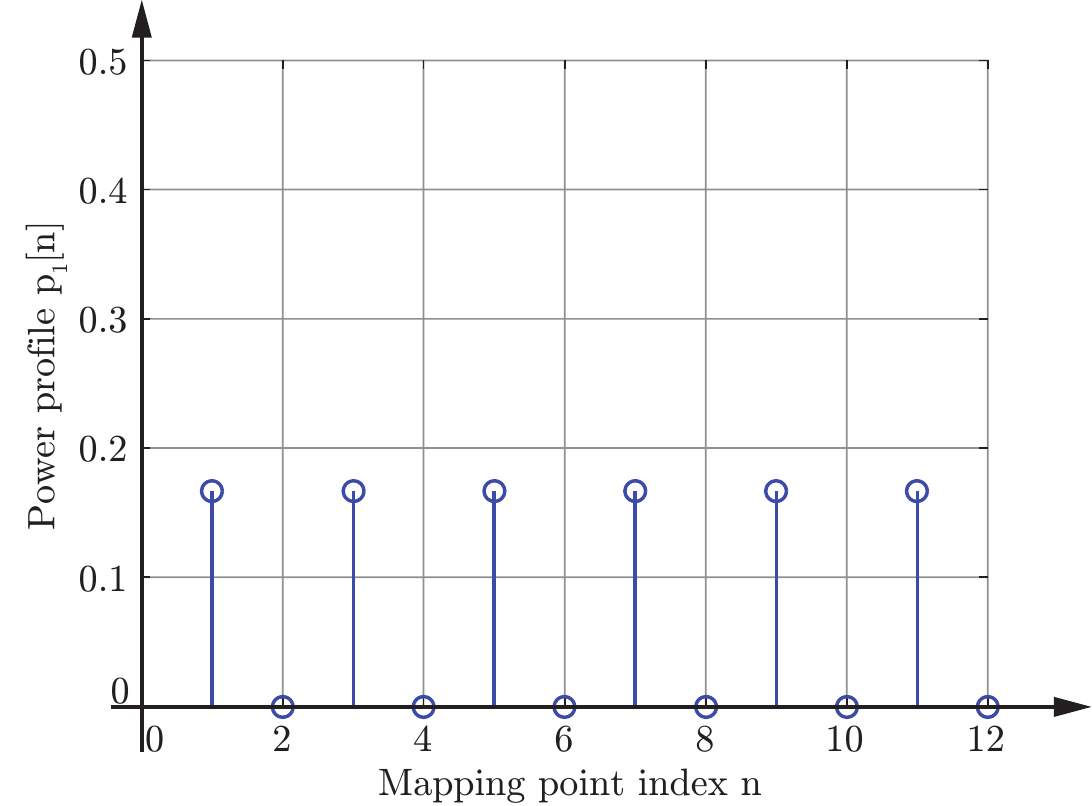}
		\label{subfig:SGN_PP_Hex}
	}
	\caption{Power profiles of circular models with one circle, i.e, $C=1$, for three stochastic interference scenarios ((a)-(c)) with $N_1 = 20$ mapping points, and for a hexagonal grid with $N_1 = 12$ mapping points, respectively.  The stochastic \ac{BS} distributions are modeled by a \ac{PPP} of intensity $\lambda = 0.01/{\rm Unit\ Area}$.  The expected number of interferers as denoted by the figure labels, is varied by altering the scenario size.}
	\label{Fig:PowerProfiles}
\end{figure}

A signal from node $\mT_{c,n}$, located at $(R_c, \Psi_{c,n})$, to a user at $(r,0)$ experiences \emph{path loss} $\ell(d_{c,n}(r))$, where $d_{c,n}(r)=\sqrt{R_c^2 + r^2 - 2R_c r \cos(\Psi_{c,n})}$ (conf. \cref{Fig:JRingScenario}) and $\ell(\cdot)$ is an arbitrary distance-dependent path loss law, as well as \emph{fading}, which is modeled by statistically independent \acp{RV} $G_{c,n}$. The received power from node $\mT_{c,n}$ at position $(r,0)$ is determined as
\begin{equation}
	P_{{\rm Rx},c,n}(r) = P_c\, p_c[n]\, \ell(d_{c,n}(r)) \, G_{c,n},
	\label{Eq:pRxSingle}
\end{equation}
where $P_c$ denotes the total transmit power of circle $c$. It is important to note that the term $P_{{\rm Rx},c,n}(r)$ can be interpreted as a \ac{RV} $G_{c,n}$, which is scaled by a factor of $P_c\, p_c[n]\, \ell(d_{c,n}(r))$.

The nodes employ omnidirectional antennas with unit antenna gain. Characteristics of antenna directivity are incorporated into the power profile. In general, the central cell will have an irregular shape that can be determined by Voronoi tesselation \citep{6515339}. For simplicity, the \emph{small ball} approximation from \citep{6515339} is applied. A user is considered as \emph{cell-edge user}, if it is located at the edge of the central Voronoi cell's inscribing ball. This approximation misses some poorly covered areas at the actual cell-edge with marginal loss of accuracy \citep{6515339}.

Let $\mS$ and $\mI$ denote the sets of nodes $\mT_{c,n}$ corresponding to desired signal and interference, respectively. Then, the aggregate signal- and interference powers are calculated as
\begin{align}
	S(r)= \sum_{\{(c,n)|\mT_{c,n}\in\mS\}}P_{{\rm Rx},c,n} (r) \label{Eq:SignalRxPower},\\
	I(r) = \sum_{\{(c,n)|\mT_{c,n}\in\mI\}}P_{{\rm Rx},c,n} (r) \label{Eq:InterfRxPower},
\end{align}
with $P_{{\rm Rx},c,n}(r)$ from \cref{Eq:pRxSingle}. The set $\mS$ may include the central node $\mT_{0,0}$ as well as nodes on the circles, if collaboration among the \acp{BS} is employed. The incoherence assumption is exploited for a more realistic assessment of the co-channel interference \citep{289422}. Following the interpretation of \cref{Eq:pRxSingle}, \cref{Eq:SignalRxPower,Eq:InterfRxPower} can be viewed as sums of scaled \acp{RV}, which are supported by a vast amount of literature for certain fading distributions such as Rayleigh, log-normal and Nakagami-m ~\citep{966578,1556824,6292935,C:Sagias_C2_2006,1096243,1673666,1583918,747812,729390,4524273,TorrieriV12,moschopoulos,1962,3717,693785,Coelho199886,abu1994outage,beaulieu1995estimating,hu2005accurate,mehta2007approximating}. 

The present work places particular focus upon the Gamma distribution due to its wide range of useful features for wireless communication engineering. The next section provides preliminary information and introduces a new theorem on the sum of Gamma \acp{RV}. The theorem is introduced before validating the accuracy of the circular model as it is later exploited for this purpose.

\section{Distribution of the Sum of Gamma \aclp{RV}}
\label{Sec:SGN_SumOfGammaTheorem}

\subsection{Preliminaries}
\label{Sec:SGN_Preliminaries}

The \ac{PDF} of a Gamma distributed \ac{RV} $X$ with \emph{shape parameter} $k$ and \emph{scale parameter} $\theta$, i.e., $X\sim\Gamma[k,\theta]$, is defined as
\begin{equation}
	f_X(x) = \frac{1}{\theta^k \Gamma(k)} x^{k-1}e^{-x/\theta},
	\label{Eq:GammaDistribution}
\end{equation}

with $k>0$ and $\theta>0$, respectively. The Gamma distribution exhibits the \emph{scaling property}, i.e., if $X\sim\Gamma[k,\theta]$, then $aX\sim[k,a\theta]$, $\forall a>0$, as well as the \emph{summation property}, i.e., if $X_i\sim\Gamma[k_i,\theta]$ with $i=1,2,\ldots,N$, then $\sum_{i=1}^N X_i\sim\Gamma[\sum_{i=1}^N k_i, \theta]$.

While the latter is convenient to apply, it is the \emph{sum of Gamma \acp{RV}} with \emph{distinct} scale parameters that has attracted a lot of attention in describing wireless communications though.
Most commonly, it emerged in the performance analysis of diversity combining receivers and the study of \emph{aggregate co-channel interference under Rayleigh fading} \citep{966578,1556824,6292935,C:Sagias_C2_2006,1096243,1673666,1583918,747812,729390,4524273,TorrieriV12}. Therefore, communication engineers have considerably pushed the search for closed form statistics. 
 
Representatively, Moschopoulos' much-cited series expansion in \citep{moschopoulos} was extended for correlated Gamma \acp{RV} in~\citep{966578}. Other approaches based on the inverse Mellin transform (e.g., \citep{mabrouk,provost}) paved the way for representations with a single integral as shown, e.g., in~\citep{6292935} or a Lauricella hypergeometric series as employed, e.g., in~\citep{1556824,1583918}.

All the aforementioned contributions focus on the sum of Gamma \acp{RV} with \emph{real-valued} shape parameter. 
The resulting integrals and \emph{infinite} series, despite being composed of elementary functions, typically yield a slow rate of convergence. Therefore, an accurate approximation by a truncated series requires to keep a high amount of terms and complicates further analysis.

The sum of Gamma \acp{RV} with \emph{integer} shape parameter has mainly been reported in statistical literature. Initial approaches focused on the moment generating function and results were obtained in the form of series expansions \citep{1962}. Based on the work of \citep{3717}, \citep{693785} was among the first to formulate a convenient closed form solution. Soon after, the \ac{GIG} distribution was published in \citep{Coelho199886}. This approach was also adopted in wireless communication engineering \citep{C:Sagias_C2_2006,1673666}. In comparison to \acp{RV} with \emph{real-valued} shape parameter, the \ac{PDF} of the sum of \acp{RV} with \emph{integer} shape parameter allows an \emph{exact} representation by a \emph{finite} series. 

\subsection{Proposed Finite Sum Representation}

In the analysis of aggregate interference statistics, it is particularly desirable to identify the main distribution-shaping factors, i.e., the interfering sources with the highest impact. However, the expressions in \citep{C:Sagias_C2_2006} and \citep{1673666} are not suitable for this task due to multiple nested sums and recursions. The proposed finite-sum representation in this work avoids recursive functions and enables to \emph{straightforwardly trace the main determinants} of the distribution characteristics.

\addtocounter{footnote}{1}
\begin{shaded}
\begin{theorem}\label{Thm:Theorem1}
Let $G_l\sim\Gamma[k_l,\theta_l]$ be $L$ independent Gamma \acp{RV} with $k_l\in\mathbb{N}^+$ and all $\theta_l$ different\footnotemark[\value{footnote}]. Then, the \ac{PDF} of $Y=G_1+\cdots+G_L$ can be expressed as
	\begin{equation}
		f_Y(y) = \sum_{l=1}^{L} \frac{\Lambda_l}{\theta_l^{k_l}}  h_{k_l-1,l}(0) e^{-y/\theta_l}
		\label{Eq:Theorem1}
	\end{equation}
	with
	\begin{align}
	\Lambda_l = \frac{(-1)^{k_l+1}}{(k_l-1)!}\prod_{i=1, i\neq l}^{L}\left(1-\frac{\theta_i}{\theta_l}\right)^{-k_i} \label{Eq:LambdaEll}, &\qquad  l=1,\ldots,L \\
	h_{\delta+1,l}(\zeta) = h_{1,l}(\zeta) h_{\delta,l}(\zeta) + \frac{d}{d\zeta} h_{\delta,l}(\zeta) \label{eq:htau}, &\qquad \delta=0,\ldots,k_l-1  
	\end{align}
	and
	\begin{align}
	h_{1,l}(0) = -y + \sum_{i=1, i\neq l}^{L} k_i \left(\frac{1}{\theta_i}-\frac{1}{\theta_l}\right)^{-1}\label{eq:h1t}, &  \qquad l=1,\ldots,L\\
	h_{1,l}^{(m)}(0) =  m! \sum_{i=1, i\neq l}^{L} k_i \left(\frac{1}{\theta_i}-\frac{1}{\theta_l}\right)^{-m - 1}\label{eq:h1taut}, &\qquad m=1,\ldots, k_{l}-1
	\end{align}
\end{theorem}
\end{shaded}
\footnotetext[\value{footnote}]{The uniqueness of $\theta_l$ can be assumed without loss of generality. In case of some $\theta_l$ being equal, the corresponding \acp{RV} are added up by virtue of the \emph{summation property of Gamma \acp{RV}} (conf. \cref{Sec:SGN_Preliminaries}).}
\begin{proof}
	The proof is provided in \cref{App:ProofOfTheorem}. 
\end{proof}

Superscript $(m)$ of $h_{1,l}^{(m)}(\zeta)$ denotes the $m$-th derivative of $h_{1,l}(\zeta)$. The recursive determination of $h_{\delta,l}(\zeta)$ in \cref{eq:htau} seemingly  interrupts the straightforward calculation of $f_Y(y)$. However, $h_{\delta,l}(\zeta)$  is a function of only $h_{1,l}(\zeta)$ and its higher order derivatives. Therefore, the function series in \cref{eq:htau} can be evaluated \emph{in advance} up to the highest required degree $\delta_{\max}= \max_l k_l-1$.

Thus, the proposed approach enables the \emph{exact} calculation of $f_Y(y)$ in a \emph{component-wise} manner
\footnote{A \emph{Mathematica\textregistered} implementation is provided at \emph{https://www.nt.tuwien.ac.at/downloads/?key=g1Y9anw3Dcletqb57RhoiH7ZleE1YlbG}. 
The code is conveniently separated into the pre-calculation, storing and reloading of the auxiliary functions in \cref{eq:htau},  and the computation of the actual distribution function.}. In the next step, it is shown how to apply \Cref{Thm:Theorem1} in the proposed circular model.

\subsection{Application in Circular Interference Model}

Assume that $G_{c,n}\sim\Gamma[k_{c,n},\theta_{c,n}]$ in \cref{Eq:pRxSingle}, with $k_{c,n}\in\mathbb{N}^+$ and $\theta_{c,n}>0$. Then, \cref{Eq:SignalRxPower,Eq:InterfRxPower} represent sums of scaled Gamma \acp{RV} $P_{{\rm Rx},c,n}(r)\sim\Gamma[k_{c,n}, \theta_{c,n}'(r)]$, where $\theta_{c,n}'(r) = P_c\, p_c[n]\, \ell(d_{c,n}(r))\,\theta_{c,n}$.
Therefore, their \acp{PDF} can be determined by applying \Cref{Thm:Theorem1}. 

The theorem requires all scale parameters to be different. Thus, let $\ftheta_{\mI}(r)$ denote the vector of unique scale parameters $\theta_{c,n}'(r)$ with $(c,n)$ from the set $\{(c,n)|\mT_{c,n} \in \mI\}$. A second vector $\fk_\mI$ contains the corresponding shape parameters. By virtue of the \emph{summation property}, if $\theta_{c,n}'(r)$ occurs multiple times in the set, the respective shape parameter in $\fk_\mI$ is calculated as the sum of shape parameters $k_{c,n}$ of the according entries. The vectors $\ftheta_\mS(r)$ and $\fk_\mS$ are obtained equivalently.
Then, the \acp{PDF} of $S(r)$ and $I(r)$ are expressed as
\begin{align}
	f_S(\gamma; r) &= \sum_{l=1}^{L_\mS} \frac{\Lambda_{l}}{\theta_{l}(r)^{k_{l}}} h_{k_{l}-1,l}(0)e^{-\gamma/\theta_{l}(r)}\label{Eq:pdfS}, \\
	f_I(\gamma; r) &= \sum_{l=1}^{L_\mI} \frac{\Lambda_{l}}{\theta_{l}(r)^{k_{l}}} h_{k_{l}-1,l}(0)e^{-\gamma/\theta_{l}(r)}\label{Eq:pdfI},
\end{align}
with $\Lambda_{l}$  and $h_{\delta,l}(\cdot)$ as defined in \cref{Eq:LambdaEll,eq:htau}. Subscript $l$ indicates the $l$-th components of the vectors $\fk_\mS$ ($\ftheta_\mS(r)$) and $\fk_\mI$ ($\ftheta_\mI(r)$) and $L_\mS$ and $L_\mI$ are their corresponding lengths, respectively. 

Hence, employing \Cref{Thm:Theorem1} allows to evaluate the \emph{exact} distributions of the aggregate signal- and interference from the circular model by \emph{finite} sums. In the following section, this fact is exploited to verify the accuracy of the model.

\section{Mapping Scheme for Stochastic Network Deployments}
\label{Sec:SGN_SamplingArbitraryNetworks}

This section presents a procedure to determine the power profiles $p_c[n]$ and the corresponding powers $P_c$ of the circular model for completely random interferer distributions. 
Then, systematic experiments are carried out to provide a reference for selecting the free variables $C$ and $N_c$, respectively. The parameters $R_c$ and $\phi_c$ are also specified by the procedure. The accuracy of the approximation is measured by means of the Kolmogorov-Smirnov distance. It is defined as
\begin{equation}
	D(r) = {\rm sup}_{x} \left| F_{I,{\rm original}}(x;r) - F_{I,{\rm circular}}(x;r)\right|,
\end{equation}
where $r$ refers to the user's eccentricity and $F_{I,{\rm original}}(x;r)$ and $F_{I,{\rm circular}}(x;r)$ denote the aggregate-interference \acp{CDF}\footnote{The \ac{CDF} of a \ac{RV} $X$ with \ac{PDF} $X$ is determined as $F_X(x) = \int_{-\infty}^x f_X(x')dx'$.} of the original deployment and the circular model, respectively. The corresponding \acp{PDF} are obtained by \Cref{Thm:Theorem1}.

\subsection{Mapping Procedure}
\label{Sec:SamplingScheme}

\begin{algorithm}
	\KwData{number of circles $C$; nodes per circle $N_c$\;
	\hspace{28pt}		original base station deployment $\mN$\;
	\hspace{28pt}		inner- and outer radius of mapping region $\mA$: $\Rin$ and $\Rout$\;}
	\KwResult{$P_c$, $p_c[n]$, $R_c$ and $\phi_c$ for all $c \leq C$\;}
	\For {$c = 1\  {\rm to}\ C$}{
          determine $R_c$ and $\phi_c$ based on the strongest interferer that has not yet been mapped\;
     }	
	\For {$c = 1\  {\rm to}\ C$}{
             specify mapping region $\mA_c$ with  inner radius $R_c$ and outer radius $R_{c+1}$\;
		\textbf{if} $c=1$ \textbf{then} set inner radius of $\mA_c$ to $\Rin$; \textbf{end} \\
		\textbf{if} $c=C$ \textbf{then} set outer radius of $\mA_c$ to $\Rout$; \textbf{end} \\
             compute $P_c$ and $p_c[n]$ for $\mA_c$\;
      }  	
	\caption{Mapping procedure for circular model.}
	\label{Alg:MappingProcedure}
\end{algorithm}

Let $\mN$ denote a (possibly heterogeneous) set of  \acp{BS}\footnote{A deployment is denoted as \emph{heterogeneous}, if the network encompasses different types of \acp{BS}. The part of a network that is associated to a certain type of \ac{BS} is denoted as \emph{tier}.} 
that are arbitrarily distributed within an annulus $\mA$ of inner radius $\Rin$ and outer radius $\Rout$, as shown in \cref{Fig:CircularModelExample}. Radius $\Rout$ as well as the number of nodes in $\mN$ could be substantially large. Given a circular model with $C$ circles and $N_c$ nodes per circle, the parameters $P_c$, $R_c$ and $\phi_c$ as well as the power profile $p_c[n]$ can be determined by \cref{Alg:MappingProcedure}.

The presented procedure employs the origin as a reference point and therefore does \emph{not} depend on the user location. 
The computation of $P_c$ and $p_c[n]$ outlines as follows. Let $\mT_{c,n}$ denote node $n$ on circle $c$. Assume that its associated mapping area $\mA_{c,n}$ is bounded by the circles of radius $R_c$ and $R_{c+1}$ (in the case of $c=1$, the inner radius is set to $\Rin$; for $c=C$ the outer radius is set to $\Rout$) as well as the perpendicular bisectors of the two line segments $\overline{\mT_{c,n}\mT_{c,n-1}}$, and $\overline{\mT_{c,n}\mT_{c,n+1}}$, as illustrated in \cref{Fig:CircularModelExample}. This yields an even division of circle $c$'s mapping area $\mA_c$, which can be formulated as $\mA_c = \bigcup_{n\in\{1,\ldots,N_c\}} \mA_{c,n}$. The average received power at the origin from all considered \acp{BS} in $\mA_c$  is calculated as 
\begin{equation}
     P_{{\rm Rx},\mA_c} = \sum_{i\, \in \, \mN\cap\mA_c}P_{{\rm Tx},i}\, \ell(d_i) \Exp[G_{i}],
    \label{Eq:AvgRxPowerccp}
\end{equation}
where $P_{{\rm Tx},i}$, $d_i$ and $G_i$ correspond to transmit power, distance and experienced fading of interferer $i$, respectively. Then, the total transmit power $P_c$ is obtained by mapping  $P_{{\rm Rx},\mA_c}$ back on the circle, which formulates as $P_c =  P_{{\rm Rx},\mA_c}\, \ell(R_c)^{-1}$. Hence, in this scheme the average received powers from the original deployment and the circular model are equivalent at the origin. The segmentation of $\mA_c$ into areas $\mA_{c,n}$ yields the corresponding power profile
\begin{equation}
	p_c[n] =\frac{1}{ P_{{\rm Rx},\mA_c}} \left(\sum_{i\, \in\, \mN\cap\mA_{c,n}} P_{{\rm Tx},i}\,  \ell(d_i)\right),
	\label{Eq:PowerProfileExSample}
\end{equation}
with $P_{{\rm Rx},\mA_c}$ from \cref{Eq:AvgRxPowerccp}. 

In the presented procedure, the parameters $R_c$ and $\phi_c$ are set such that the $c$-th dominant interferer coincides with a node on circle $c$, as illustrated in \cref{Fig:CircularModelExample}. This ensures that $R_1\geq \Rin$ (in a heterogeneous network, as investigated in \cref{Sec:MappingPerformanceHetNet}, non-dominant interferers between $\Rin$ and $R_1$ are mapped "back" on circle $1$ by the receive-power dependent weighting in \cref{Eq:PowerProfileExSample}) and $R_C\leq \Rout$, and is especially suitable for completely random interferer distributions, as demonstrated in the next section. In fully regular scenarios, on the other hand, a circle comprises multiple, equally dominant nodes, making it expedient to specify $R_c$ and $\phi_c$ according to the structure of the grid. For example, the circular model allows to \emph{perfectly} represent a hexagonal grid setup, when the number of mapping points is set as a multiple of six. Then, the nodes on the circle coincide with the actual interferer locations. An  exemplary power profile for $N_1 = 12$ is shown in \cref{subfig:SGN_PP_Hex}.

\cref{Alg:MappingProcedure} is one of many possible mapping approaches. It is a heuristic, based on the authors' experience and observations and is thus \emph{not claimed} to be optimal and its refinement yields an interesting topic for further work. The next two sections perform systematic experiments in \emph{completely random} scenarios to provide a reference for setting $C$ and $N_c$. For reasons of clarity, \cref{Sec:SamplingPerformanceEval} is limited to \emph{homogeneous} \ac{BS} deployments. Heterogeneous setups are then evaluated in \cref{Sec:MappingPerformanceHetNet}. 
It is refrained from stochastic scenarios with a certain degree of regularity, since measuring spatial inhomogeneity is an ongoing topic of research~\citep{5621983}. Completely random- and fully regular scenarios are considered as limiting cases, encompassing every conceivable practical deployment.

\subsection{Performance Evaluation for Homogeneous Base Station Deployments}
\label{Sec:SamplingPerformanceEval}

\begin{table}
\centering
	\caption{System setup for evaluation.}
\begin{tabular}{ r | l }
	\hline
	\textbf{Parameter} & \textbf{Value} \\
	\hline
      Transmit power & $P_{T1} = 1$ ($P_{T2} = 0.01$) \\
	Node density & $\lambda = \{0.05, 0.1\}/{\rm Unit\  Area}$ ($\lambda_2 = 1/{\rm Unit\  Area}$) \\
	Expected number of interferers & $\NI=\{100,1000\}$ \\
	\hline 
	Antenna configuration & $N_{\rm Tx}\times N_{\rm Rx}=2\times 1$; omnidirectional\\
	Path loss & $\ell(x)=\max(c_{\rm B}, c_{\rm PL}\, x^{-4})$, $c_{\rm B}=1$, $c_{\rm PL} = 1,x> 0$\\
	Fading & $G_{c,n}\sim\Gamma[2,1]$\\
	\hline
\end{tabular}
	\label{Tab:ParameterSetup}
\end{table}

The \emph{original} interferer deployment $\mN$ is modeled by a \ac{PPP} of intensity $\lambda$. Such process is considered \emph{most challenging} for the regularly structured circular model, as it represents complete spatial randomness. Signal attenuation is modeled by a log-distance dependent path loss law $\ell(x) = \max(c_{\rm B}, c_{\rm PL}\, x^{-4})$, and Gamma fading with $k=2$ and $\theta=1$, referring to a $2\times1$ \ac{MISO} setup and maximum ratio transmission. Without loss of generality, normalized distance values $x$ are used. The dimension of the network is incorporated in the intercept $c_{\rm B}$ and the constant $c_{\rm PL}$, respectively.  In this work, $c_{\rm B}=1$ and $c_{\rm PL}=1$ for simplicity\footnote{Consider the examples as presented in normalized setups, i.e., relative to multiples of the wavelength.}. The \acp{BS} transmit with unit power $P_{T1}=1$ and are distributed within an annular regions of inner radius $\Rin = 2$ and $\Rout=\sqrt{N_{\rm I}/(\pi \lambda) +\Rin^2}$. Radius $\Rin$ ensures a unit central cell size, assuming that the central \ac{BS} also transmits with $P_{T1}$. The outer radii $\Rout$ are chosen such that, on average, $\NI$ \acp{BS} locations are generated within the corresponding annulus\footnote{Consider a \ac{PPP} of intensity $\lambda$ within an annulus of inner radius $\Rin$ and outer radius $\Rout$. The expected number of generated nodes is calculated as $\NI = \lambda(R_{\rm out}^2 - R_{\rm in}^2)\pi$.}. In order to cover a wide range of scenarios, $\NI = \{100,1000\}$ and $\lambda=\{0.1, 0.05\}/{\rm Unit\  Area}$ are studied. The parameter settings are summarized in \cref{Tab:ParameterSetup}.

\begin{figure}
	\centering
		\subfloat[$N_c = 10$, $\lambda = 0.1/{\rm Unit\ Area}$]{\includegraphics[width = \mydoubleW]{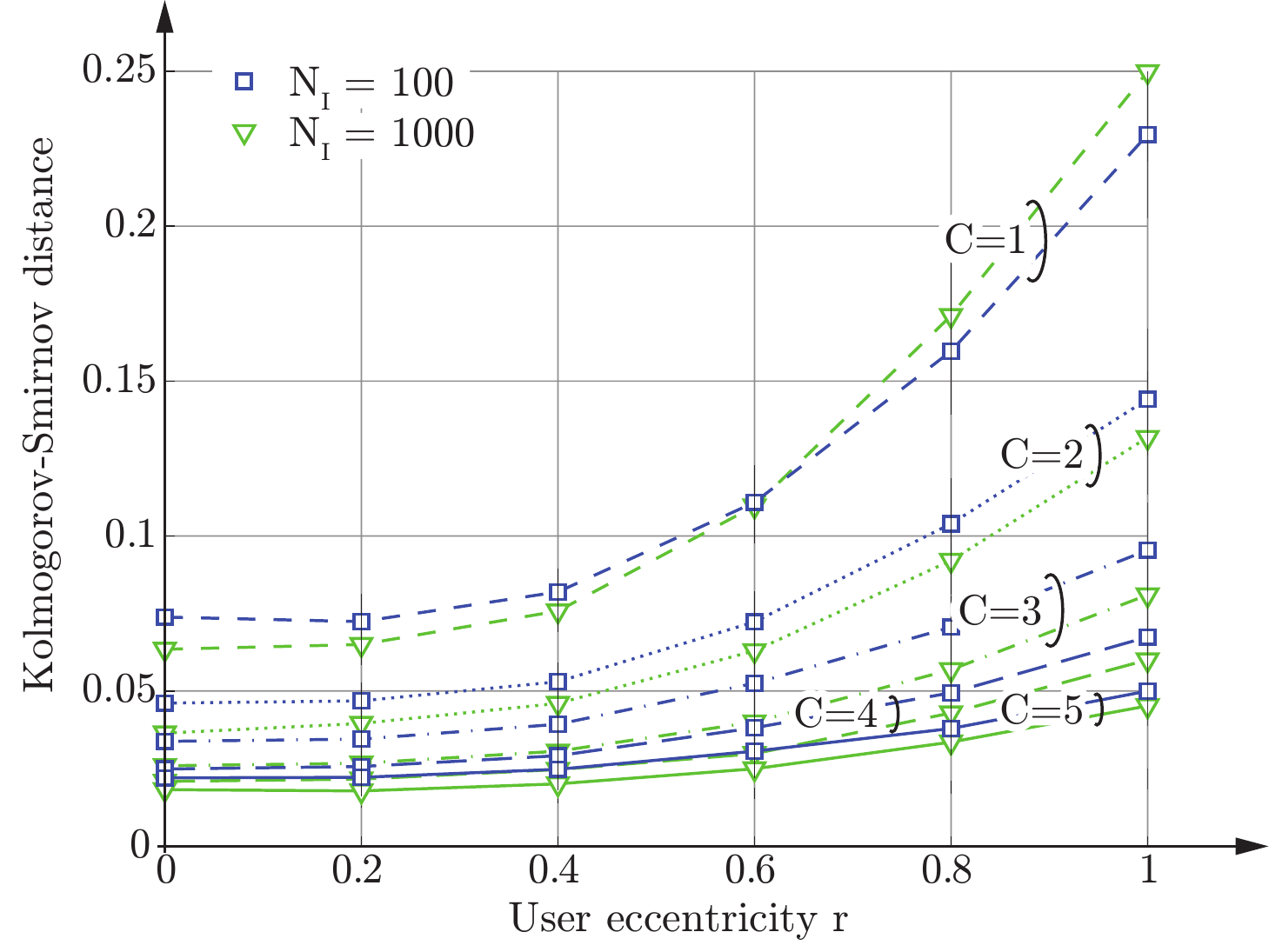}
		\label{subfig:SGN_KS_distance_10_01}}
		\subfloat[$N_c = 20$, $\lambda = 0.1/{\rm Unit\ Area}$]{\includegraphics[width = \mydoubleW]{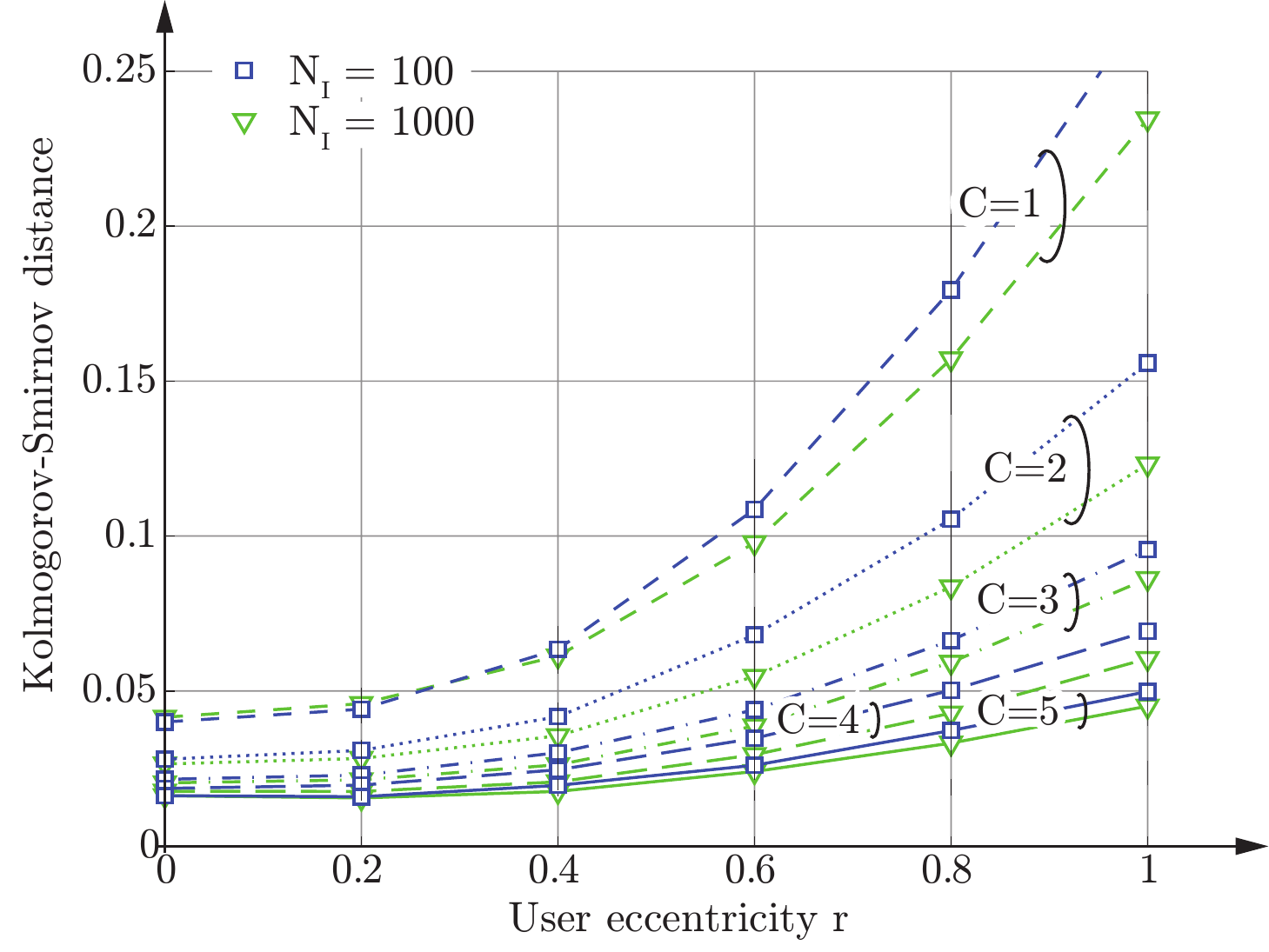}
		\label{subfig:SGN_KS_distance_20_01}}\\
		\subfloat[$N_c = 20$, $\lambda = 0.05/{\rm Unit\ Area}$]{\includegraphics[width = \mydoubleW]{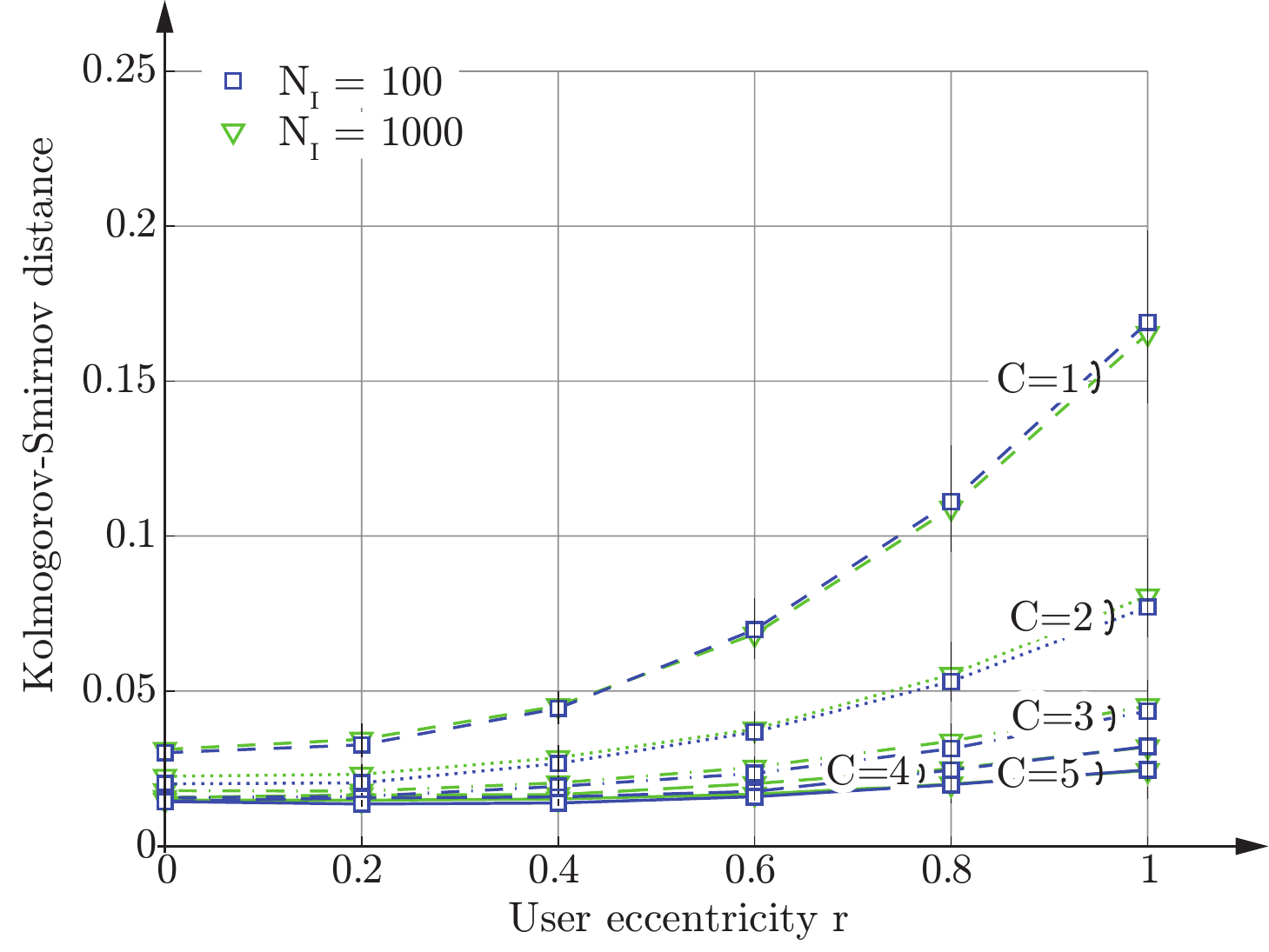}
		\label{subfig:SGN_KS_distance_20_005}}
		\subfloat[$N_c = 40$, $\lambda = 0.05/{\rm Unit\ Area}$]{\includegraphics[width = \mydoubleW]{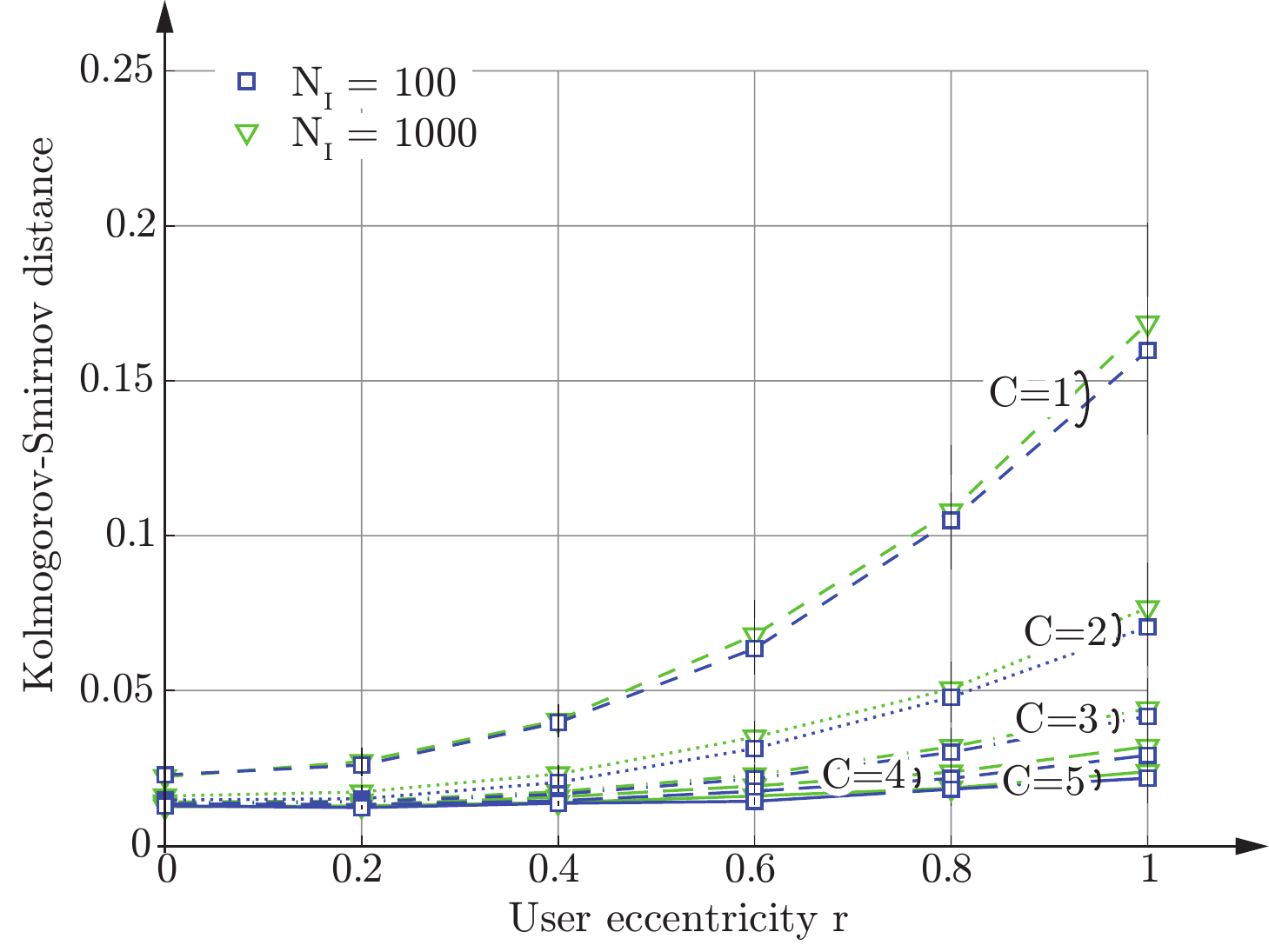}
		\label{subfig:SGN_KS_distance_40_005}}\\
	\caption{Kolmogorov-Smirnov distance over user eccentricity $r$. Plot markers $\{\bigtriangledown,\Box\}$ refer to various scenario sizes $\NI=\{100,1\,000\}$ expected interferers, respectively. Different line styles denote circular models with $C = \{1,2,3,4,5\}$. Figure labels refer to the corresponding number of nodes per circle, $N_c$, and the density $\lambda$ of the original interferer deployment. Black bars depict $95\%$ confidence intervals.}
	\label{Fig:KS_distances_over_r}
\end{figure}

For each scenario snapshot, ten circular models with $C = \{1,2,3,4,5\}$ and two distinct values of $N_c$ are set up according to \cref{Sec:SamplingScheme}. In the case of  $\lambda=0.1/{\rm Unit\  Area}$, $N_c=\{10,20\}$ and, for $\lambda = 0.05/{\rm Unit\  Area}$, $N_c=\{20,40\}$, respectively. Then, the \emph{aggregate interference distributions} are determined. The distributions for the original interferer deployment are only obtained via simulations (by averaging over $1\,000$ spatial realizations and $10\,000$ fading realizations), since the vast amount of nodes hampers the application of \Cref{Thm:Theorem1} due to complexity issues. On the other hand, the circular models comprise at most $44$ \emph{active} nodes and therefore enable to utilize the theorem. This number is obtained for $C=5$ and $N_c=40$, and stems from the fact that in a homogeneous \ac{BS} deployment, the dominant interferers are also the closest ones. Therefore, the presented scheme only maps a single \ac{BS} on each circle $c<C$, i.e., except for $c=C$ there is only one active node per circle. 

\cref{Fig:KS_distances_over_r} depicts Kolmogorov-Smirnov distances over the user eccentricity $r$. The first important observation is that the accuracy considerably improves with an increasing number of circles $C$. This mainly results from accurately capturing the first few dominant \acp{BS} that have the largest impact on the aggregate interference distribution, as later shown in \cref{Sec:SGN_GenericScenario}. A second remarkable observation is that doubling the amount of nodes per circle from $N_c = 10$ to $N_c=20$ for $\lambda = 0.1/{\rm Unit\ Area}$ (conf. \cref{subfig:SGN_KS_distance_10_01,subfig:SGN_KS_distance_20_01}), and from $N_c = 20$ to $N_c=40$ for $\lambda = 0.05/{\rm Unit\ Area}$ (conf. \cref{subfig:SGN_KS_distance_20_005,subfig:SGN_KS_distance_40_005})  does not achieve smaller Kolmogorov-Smirnov distances, respectively. This result indicates that it is rather the number of circles $C$ than the number of nodes per circle $N_c$ that impacts the accuracy. As shown in the examples, good operating points are $N_c = \lfloor 1/\lambda\rfloor$ and $C={\rm arg\, min}_c |\ell(d_c)/\ell(d_1) - 2\,\lambda|$, where $d_c$ denotes the average distance of the $c$-th dominant interferer to the origin~\citep{moltchanov2012distance}. Lastly, it should be noted that the circular model allows to represent $1\,000$ and more interferers by some $10$ nodes with Kolmogorov-Smirnov distances at the cell-edge not exceeding $0.05$.

\subsection{Performance Evaluation of Heterogeneous Base Station Deployments}
\label{Sec:MappingPerformanceHetNet}
\begin{figure}
	\centering
		\subfloat[$N_c = 10$]{\includegraphics[width = \mydoubleW]{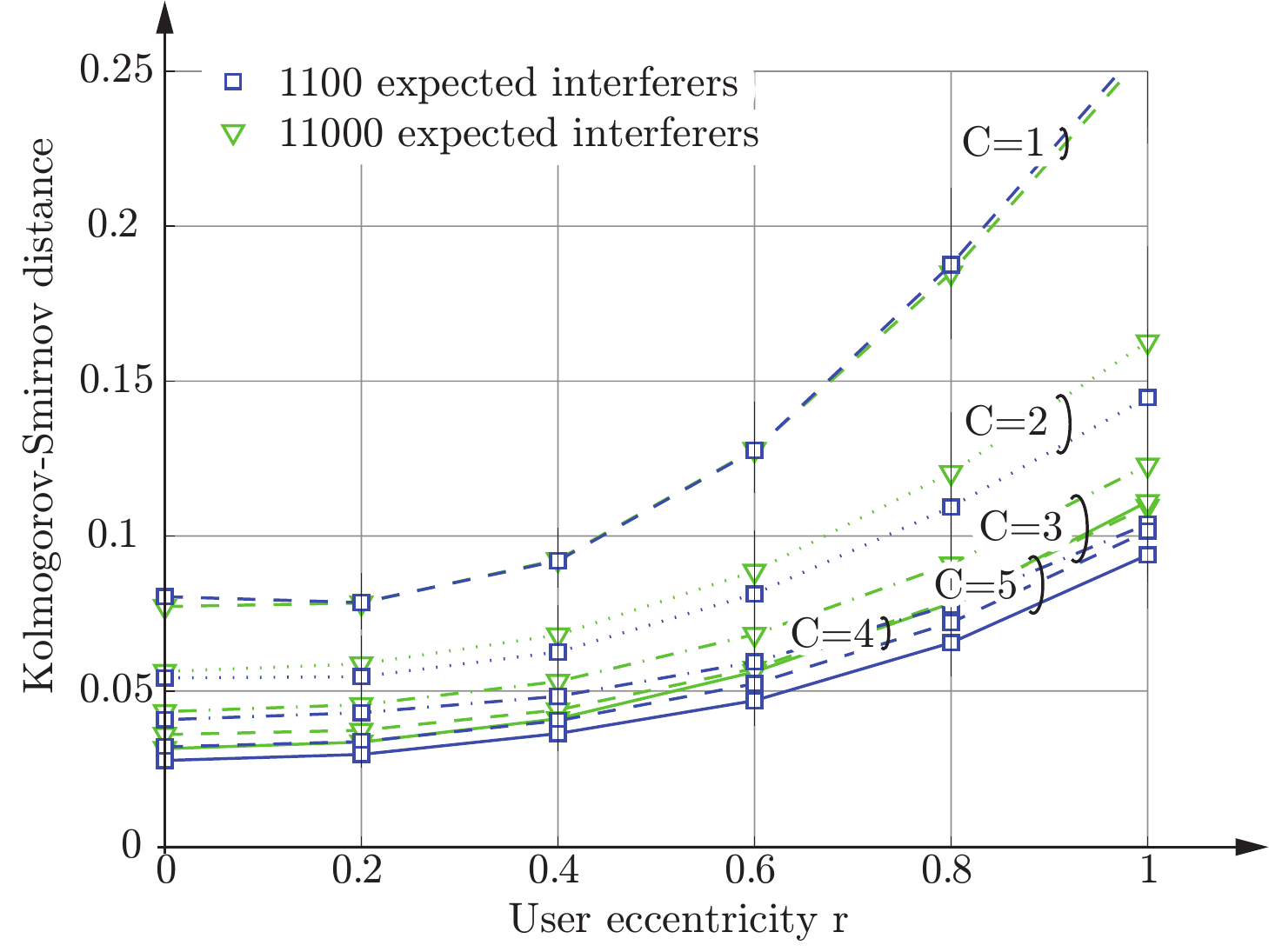}
		\label{subfig:SGN_KS_distance_10_1}}
		\subfloat[$N_c = 20$]{\includegraphics[width = \mydoubleW]{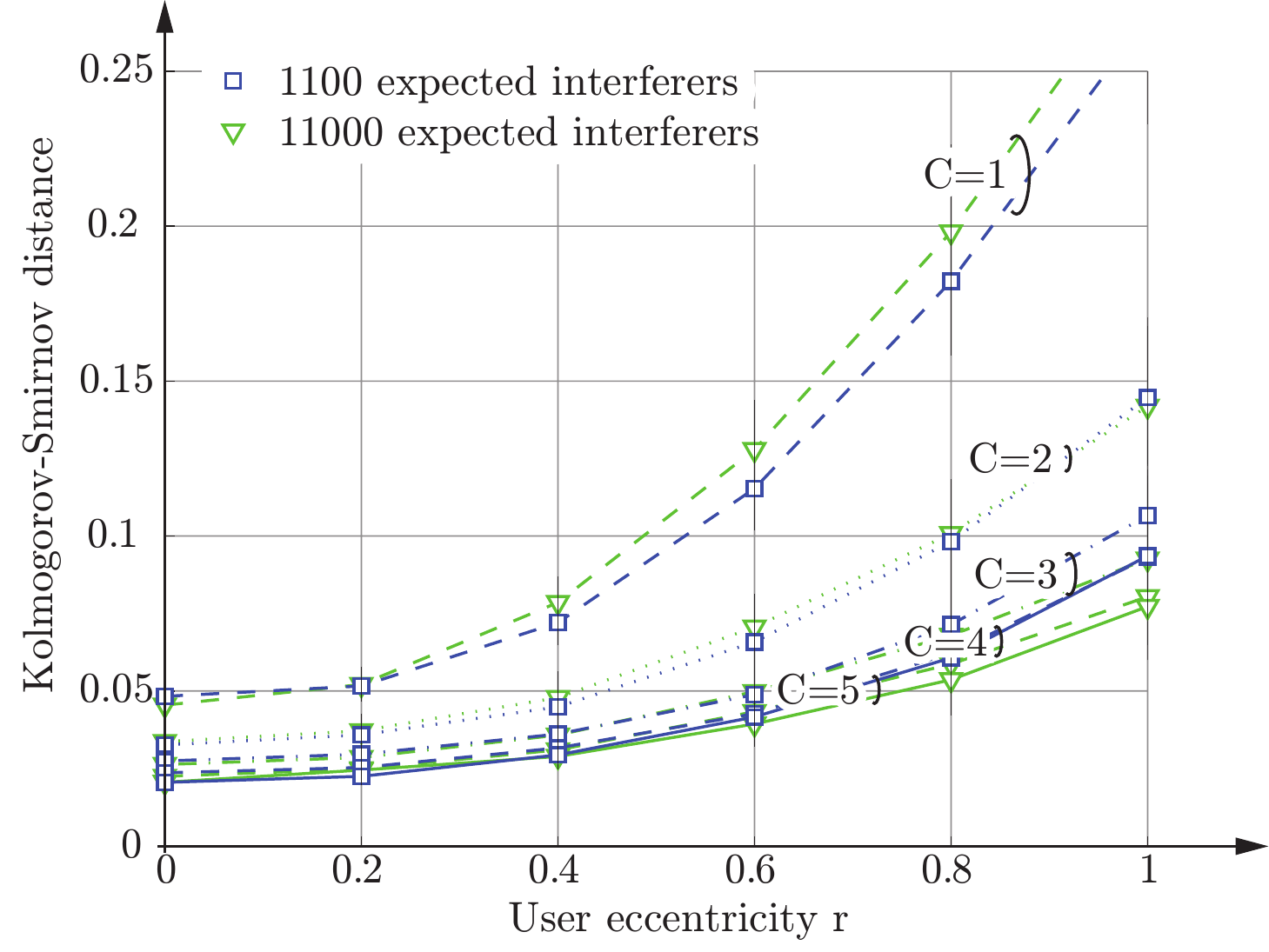}
		\label{subfig:SGN_KS_distance_20_1}}
	\caption{Kolmogorov-Smirnov distance over user eccentricity $r$ for \emph{heterogeneous} \ac{PPP} scenarios with $\lambda =0.1/{\rm Unit Area}$ ($P_{T1}=1$) and $\lambda_2 = 1/{\rm Unit\  Area}$ ($P_{T2}=0.01$). Plot markers $\{\bigtriangledown,\Box\}$ refer to various scenario sizes with $\{1\,100,11\,000\}$ expected interferers, respectively. Different line styles denote circular models with $C = \{1,2,3,4,5\}$. Figure labels refer to the corresponding number of nodes per circle $N_c$. Black bars depict $95\%$ confidence intervals. }
	\label{Fig:KS_distances_over_r_het}
\end{figure}

In this section, a second independent \ac{PPP} of intensity $\lambda_2 = 1/{\rm Unit\  Area}$ is added on top of the \ac{PPP} scenarios with $\lambda=0.1$ in \cref{Sec:SamplingPerformanceEval}. The corresponding nodes transmit with normalized power $P_{T2} = 0.01$, thus representing a dense overlay of low power \acp{BS}. For simplicity, they are distributed within annuli of inner radius $\Rin$ and outer radii $\Rout$ as specified above\footnote{To ensure a unit central cell size, an inner radius of $1+(P_{T1}/P_{T2})^{-1/\alpha}$ would be sufficient.}. Then, the total number of expected interferers calculates as $\{1\,100,11\,000\}$, respectively. For each snapshot, \cref{Alg:MappingProcedure} is applied with $C=\{1,2,3,4,5\}$ and $N_c = \{10,20\}$. The performance evaluation is carried out along the lines of \cref{Sec:SamplingPerformanceEval} and the parameters are summarized in \cref{Tab:ParameterSetup}.

\cref{Fig:KS_distances_over_r_het} depicts the results in terms of Kolmogorov Smirnov distances. It is observed that employing the same parameters $C$ and $N_c$ as for the homogeneous scenarios only slightly decreases the performance (conf. \cref{subfig:SGN_KS_distance_10_01,subfig:SGN_KS_distance_20_01}), although mapping $11$ times as many interferers. 
Hence, applying the recommendations in \cref{Sec:SamplingPerformanceEval} with respect to the \ac{PPP} that models the \acp{BS} with the highest power, yields a good initial operating point.

\subsection{Power Profiles of \ac{PPP} Snapshots}
\label{Sec:InterferenceCharacteristicsPPP}

\begin{figure}
	\centering
	\includegraphics[width=.55\columnwidth]{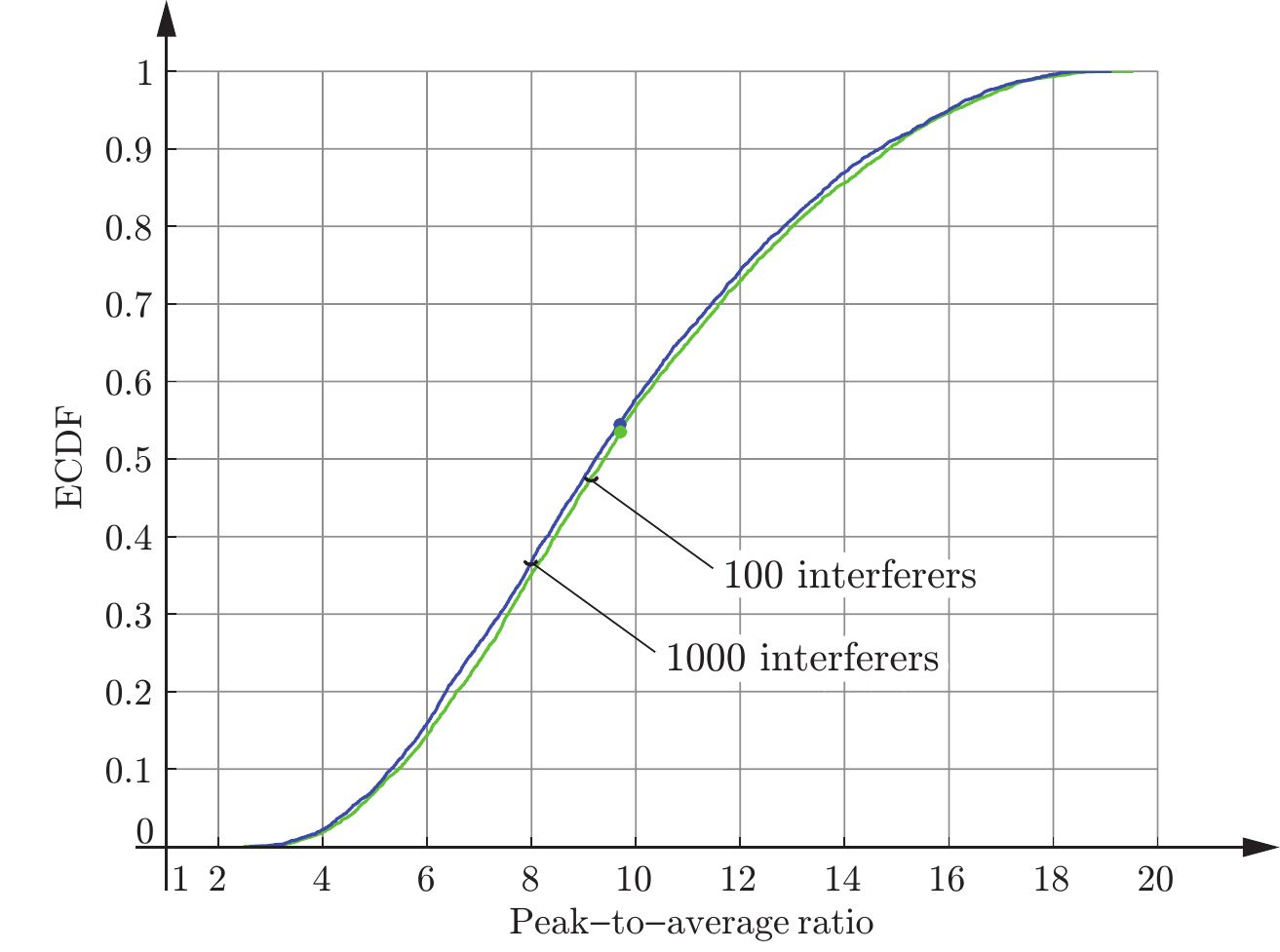}
	\caption{Peak-to-average ratio of power profiles of \ac{PPP} scenarios with intensity $\lambda = 0.1/{\rm Unit\  Area}$ and $\NI = \{100,1000\}$ expected interferers. The corresponding circular models are obtained by \cref{Alg:MappingProcedure} with $C=1$ and $N_1 = 20$. Bold dots denote the mean ratios.}
	\label{Fig:SGN_PAPR_ECDF}
\end{figure}

As indicated in \cref{Fig:PowerProfiles}, power profiles of homogeneous \ac{PPP} scenarios are characterized by one or a few large amplitudes. To quantify this claim, \cref{Fig:SGN_PAPR_ECDF} shows the empirical distributions of the power-profile peak-to-average ratios as obtained from the \acp{PPP} in \cref{Sec:SamplingPerformanceEval} with $\lambda=0.1/{\rm Unit\  Area}$. The corresponding circular models encompass a single circle (i.e., $C=1$) with $N_1 = 20$ mapping points. It is observed that the peak-to-average ratios range from $3$ to $19$ with the medians being located around $9.5$. The presence of dominant interferers results in a large asymmetry of the interference impact. However, in modeling approaches that are based on stochastic geometry, the differences between scenarios at both ends of the scale are concealed by spatial averaging. What is more, such approaches commonly require user-centric isotropy of the setup in order to obtain exact solutions (e.g., circularly symmetric exclusion regions~\citep{6515339}). Hence, the differences between interference characteristics in the \emph{center of the cell} and at \emph{cell-edge} are generally not accessible. The next section applies the circular model to generate a generic, circularly symmetric scenario and, by employing \Cref{Thm:Theorem1}, analyzes the impact of user eccentricity.

\section{Interference and Rate at Eccentric User Locations}
\label{Sec:SGN_GenericScenario}

This section investigates user-centric \ac{BS} collaboration schemes in scenarios with \emph{asymmetric interferer impact}. The asymmetry can either arise from \emph{non-uniform} power profiles or user locations outside the center of an otherwise isotropic scenario. The particular emphasis of this section is on the latter, since it is found less frequently in literature. In order to generate a generic, circularly symmetric scenario\footnote{In fact, the circular model generates a \emph{rotationally symmetric} scenario due to the finite number of nodes. However, by setting $N_c$ sufficiently large, the scenario can be considered as \emph{quasi-circularly symmetric}.}, the introduced circular model is applied, which enables to employ \Cref{Thm:Theorem1} for the analysis of the interference statistics.





\subsection{Generic Circularly Symmetric Scenario}
\label{Sec:GenericCircularModel}

\begin{figure}
	\centering
	\includegraphics[width=.55\textwidth]{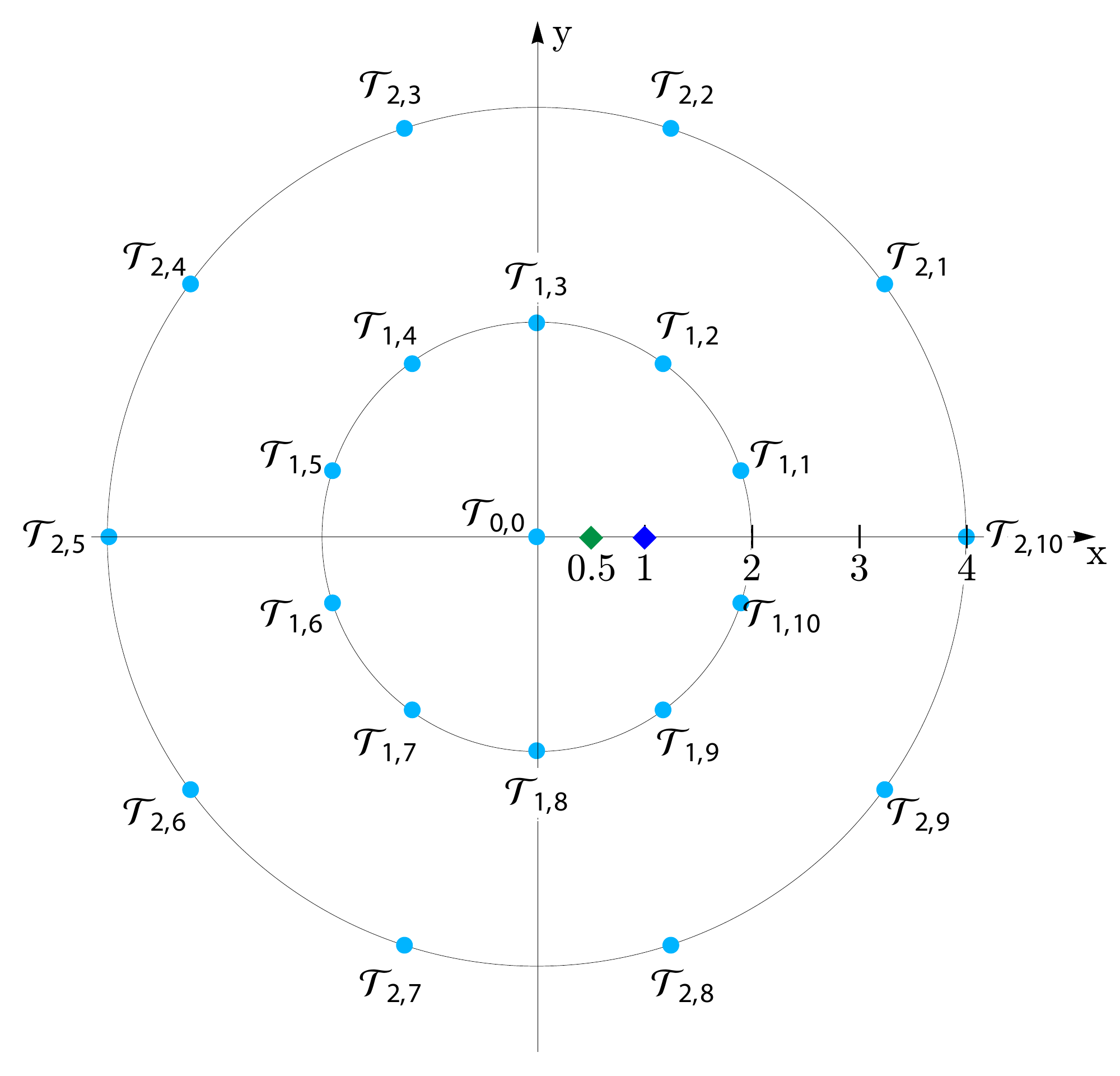}
	\caption{Circular model with two two circles of radius $R_1 = 2$ and $R_2=4$, respectively. Each circle employs $10$ transmitters. The transmitter positions are rotated by $\phi_1 = -\pi/10$ and $\phi_2=0$. Users at $r=0.5$ and $r=1$ are denoted as bold dots and refer to \emph{middle of cell} and \emph{cell-edge}, respectively.}
	\label{Fig:AppScenario}
\end{figure}

The network is composed of a central \ac{BS} and two circles of interferers with $R_1=2$ and $R_2 = 4$, as depicted in \cref{Fig:AppScenario}. Each circle employs $10$ interferers, a uniform power profile, i.e., $p_c[n]=1/10$, and unit total transmit power, i.e., $P_c= 1$. The interferer locations are assumed to be rotated by $\phi_1 = -\pi/10$ and $\phi_2=0$, respectively. \ac{BS} $\mT_{0,0}$ is located at the origin and $P_0=0.1$. The normalized system parameters are employed to emulate a unit central cell size and to facilitate reproducibility.

The parameters of the circular model are summarized in \cref{Tab:ParamsOfCirc} and the modeling of the signal propagation is referred from \cref{Tab:ParameterSetup}, respectively. The first goal is to identify the nodes, which dominate the interference statistics at eccentric user locations. Then, these insights are applied for user-centric \ac{BS} coordination and -cooperation.

\subsection{Components of Asymmetric Interference}
\label{Sec:InterferenceComponents}

\begin{table}
\centering
	\caption{Parameters of circular model for numerical evaluation.}
\begin{tabular}{c | l  l  l  l  l  l}
	\hline
	\textbf{Circle} & \multicolumn{5}{|l}{\textbf{Values}} \\
	\hline
	 1 & $R_1 = 2$ & $N_1 = 10$ & $P_1 = 1$ &  $\phi_1 = -\frac{\pi}{10}$ & $p_1[n] = \frac{1}{10}$ & $n\in\{1,\ldots,10\}$\\
	 2 & $R_2 = 4$ & $N_2 = 10$ & $P_2 = 1$ & $\phi_2 = 0$ & $p_2[n] = \frac{1}{10}$ &  $n\in\{1,\ldots,10\}$ \\
	\hline
\end{tabular}
	\label{Tab:ParamsOfCirc}
\end{table}

In the first step, only the inner circle of interferers is assumed to be present, i.e., the set $\mI$ comprises the $10$ nodes $\mT_{1,n},\, n=1,\ldots,10$, of circle $1$. The target is to determine the impact of the closest nodes on the aggregate interference statistics. For this purpose, two representative user locations at $r=R_1/4$ and $r=R_1/2 $ are investigated, referring to \emph{middle of cell} and  \emph{cell-edge}, respectively.

The \ac{PDF} of the aggregate interference is obtained by \Cref{Thm:Theorem1}. Its evaluation is simplified by the scenario's symmetry about the $x$-axis: (i) equal node-to-user distances from upper- and lower semicircle, i.e., $d_{1,n} = d_{1,10-n+1}$, (ii) uniform power profile $p_1(n) = 1/10$, and (iii) equal scale parameters $\theta_{1,n} = 1$. Thus, $\theta'_{1,n}(r)=\theta'_{1,10-n+1}(r)$, with  $\theta'_{1n}(r) = P_1/10\,  \ell(d_{1,n}(r))$. The vectors $\ftheta_\mI(r)$ and $\fk_\mI$ 
are of length $L^\mI=5$, with $\left[\ftheta_\mI(r)\right]_l = \theta'_{1,l}(r)$ and $\left[\fk_\mI\right]_l = 4$, respectively. Hence, the distribution of aggregate interference at distance $r$ from the center formulates as
\begin{equation}
	f_I(x;r) = \sum_{l=1}^5 \frac{\Lambda_l}{\theta'_{1,l}(r)^4} h_{3,l}(0) e^{-x/\theta'_{1,l}(r)},
	\label{Eq:interferenceDistroOneRing}
\end{equation}
where
\begin{align}
	\Lambda_l &= -\frac{1}{6}\prod_{i=1, i\neq l}^{5}\left(1-\frac{\theta_i}{\theta_l}\right)^{-4} \label{Eq:LambdaEllEx}, \qquad  l=1,\ldots,5, \\
	h_{3,l} &= \left(h_{1,l}(0)\right)^3 + 3h_{1,l}(0) h_{1,l}^{(1)}(0) + h_{1,l}^{(2)}(0),
\end{align}
with
\begin{align}
	h_{1,l}(0) &= -y + 4 \sum_{i=1, i\neq l}^{5} \left(\frac{1}{\theta_i}-\frac{1}{\theta_l}\right)^{-1}\label{eq:h1tEx},\\
	h_{1,l}^{(1)}(0) &= 4 \sum_{i=1, i\neq l}^{5}  \left(\frac{1}{\theta_i}-\frac{1}{\theta_l}\right)^{-2}, \\
	h_{1,l}^{(2)}(0) &=  8 \sum_{i=1, i\neq l}^{5}  \left(\frac{1}{\theta_i}-\frac{1}{\theta_l}\right)^{-3}.
\end{align}

\begin{figure}
	\centering
	\includegraphics[width=\customwidthsmall]{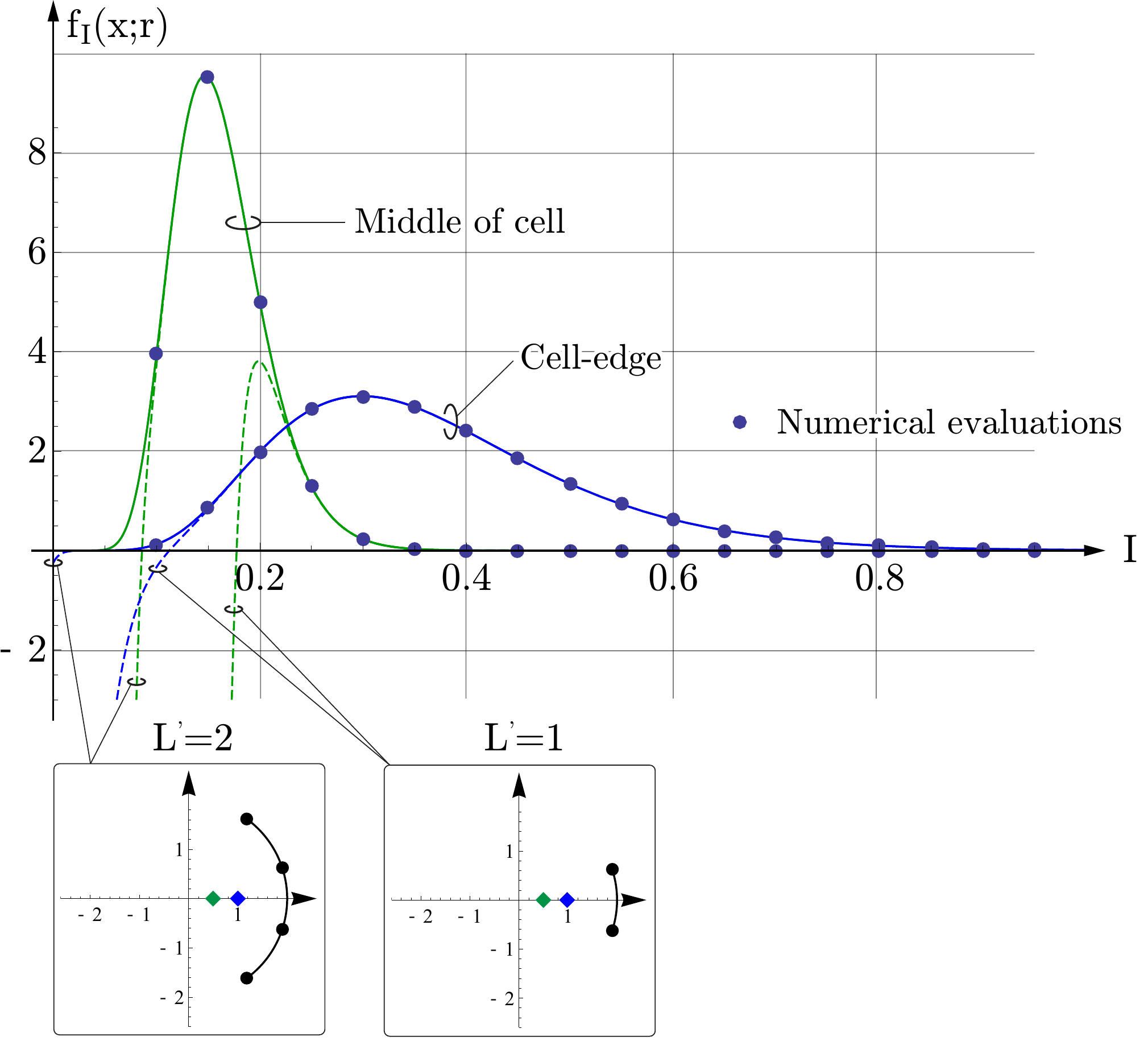}
	\caption{Distribution of aggregate interference at user distances $r=0.5$ and $r=1$, respectively. Dots refer to results as obtained with the approach in \citep{6292935}. Dashed curves show contribution from dominant interferers.}
	\label{Fig:InterferenceComponents}
\end{figure}

\cref{Fig:InterferenceComponents} shows $f_I(x;r)$ for $r=0.5$ (narrow solid curve) and $r=1$ (wide solid curve), referring to middle of cell and cell-edge, respectively. The dots denote results as obtained with the approach in \citep{6292935}, which requires numerical evaluation of a line-integral and confirms the accuracy of the proposed finite-sum representation.

In \cref{Eq:interferenceDistroOneRing}, each sum term refers to a pair of transmitters $\{\mT_{1,l},\, \mT_{1,10-l+1}\}$. The contribution of each pair to the final \ac{PDF} is rendered visible by truncating the sum in \cref{Eq:interferenceDistroOneRing} at $L'$ with $L'\in\{1,\ldots,5\}$, i.e., only the first $L'$ sum terms are taken into account. Dashed curves in \cref{Fig:InterferenceComponents} depict results for $L'=1$ and $L'=2$.

It is observed that (i) in the \emph{middle of the cell}, body and tail of the \ac{PDF} are mainly shaped by the \emph{four closest} interferers while (ii) at \emph{cell-edge} the distribution is largely dominated by the \emph{two closest} interferers, and (iii) interference at $r=1$ yields a larger variance than at $r=0.5$ due to higher diversity of the transmitter-to-user distances. The results verify link-level simulations in \citep{PDL06:SCVT}. They emphasize the strong impact of interference asymmetry due to an eccentric user location, which is commonly overlooked in stochastic geometry analysis. The next section exploits the above findings for \ac{BS} coordination and -cooperation and investigates the resulting \ac{SIR}- and rate statistics.

\subsection{Transmitter Collaboration Schemes}
\label{Sec:SGN_TransmitterCollabSchemes}

This subsection studies \ac{SIR}- and rate statistics in the full two-circle scenario, as shown in \cref{Fig:AppScenario}. Motivated by the observations in \cref{Sec:InterferenceComponents}, three schemes of \emph{\ac{BS} collaboration} are discussed:
\begin{enumerate}
\item \textit{No collaboration among nodes}: This scenario represents the \emph{baseline}, where $\mS = \{\mT_{0,0}\}$ and $\mI$ comprises all nodes on the circle, i.e., $\mI=\{\mT_{c,n}\}$ with $c=1,2$ and $n\in\{1,\ldots,10\}$.
\item \textit{Interference coordination}\footnote{Conf., e.g., \ac{eICIC} in the \acs{3GPP} \acs{LTE} standard \citep{3gpp.36.839}.}: The nodes coordinate such that co-channel interference from the two strongest interferers of the inner circle, $\mT_{1,1}$ and $\mT_{1,10}$, is eliminated. This could be achieved, e.g., by joint scheduling. Then, $\mS=\{\mT_{0,0}\}$ and $\mI$ is composed of $\{\mT_{1,n}\}$ with $n\in\{2,\ldots,9\}$ and $\{\mT_{2,n}\}$ with $n\in\{1,\ldots,10\}$.
\item \textit{Transmitter cooperation}\footnote{Conf., e.g., \ac{CoMP} in the \acs{3GPP} \acs{LTE} standard \citep{3gpp.36.819}.}: The signals from the two closest nodes of the inner circle, $\mT_{1,1}$ and $\mT_{1,10}$, can be exploited as useful signals and are incoherently combined with the signal from $\mT_{0,0}$. Then, $\mS=\{\mT_{0,0}, \mT_{1,1}, \mT_{1,10}\}$ and, as above, $\mI$ comprises $\{\mT_{1,n}\}$ with $n\in\{2,\ldots,9\}$ and $\{\mT_{2,n}\}$ with $n\in\{1,\ldots,10\}$.
\end{enumerate}

For each \emph{collaboration} scheme, the \acp{PDF} of aggregate signal and -interference, $f_S(x;r)$ and $f_I(x;r)$, are calculated using \Cref{Thm:Theorem1}. 
The \emph{\ac{SIR}} at user location $(r,0)$ is defined as $\gamma(r) = {S(r)}/{I(r)}$. According to \citep{1941}, the \ac{PDF} of $\gamma(r)$ is calculated as
\begin{equation}
	f_\gamma(\gamma; r) = \int_0^{\infty} z f_S(z\, \gamma; r) f_I(z;r) dz,
	\label{Eq:QuotientDistribution}
\end{equation}
where $z$ is an auxiliary variable, $f_S(\cdot;r)$ and $f_I(\cdot; r)$ refer to \cref{Eq:pdfS,Eq:pdfI}, and the integration bounds are obtained by exploiting the fact that $f_S(\gamma; r) = 0$ and $f_I(\gamma;r) = 0$ for $x< 0$.

Evaluating \cref{Eq:pdfS,Eq:pdfI} yields sums of elementary functions of the form $a \gamma^b e^{-c\gamma}$, with the generic parameters $a\in\mathbb{R}$, $b\in\mathbb{N}^+$ and $c>0$. Therefore, $f_S(\gamma;r)$ and $f_I(\gamma;r)$ can generically be written as
\begin{align}
	f_S(\gamma;r) &= \sum_{s}a_s \gamma^{b_s}e^{-c_s \gamma},\\
      f_I(\gamma;r) &= \sum_{i}a_i \gamma^{b_i}e^{-c_i \gamma},
\end{align}
and allow to straightforwardly evaluate \cref{Eq:QuotientDistribution} as
\begin{align}
	f_{\gamma}(\gamma;r) &= \sum_s\sum_i \int_0^\infty z\,  a_s (z\gamma)^{b_s} e^{-c_s(\gamma z)}\,  a_i z^{b_i}e^{-c_i z} dz  \nonumber\\ 
	&=\sum_s\sum_i a_s a_i \gamma^{b_s}(c_i+c_s \gamma)^{-i-b_s-b_i}\Gamma(i+b_s+b_i).
	\label{Eq:GenericSIR}
\end{align}

The \emph{normalized rate} $\tau$ as a function of the \ac{SIR} $\gamma(r)$ is calculated by the well known Shannon formula $\tau(\gamma(r)) = \log_2 \left(1+\gamma(r)\right)$.
Note that $\tau(\cdot)$ is a function of the \ac{RV} $\gamma(r)$. Hence, its distribution is calculated by the transformation
\begin{align}
	f_{\tau}(\tau;r) = \log_{e}(2) 2^\tau f_\gamma(2^\tau-1;r),
	\label{Eq:RatePDF}
\end{align}
with $f_{\gamma}(\cdot;\cdot)$ from \cref{Eq:GenericSIR}. 

\begin{figure}
	\centering
	\includegraphics[width=.65\textwidth]{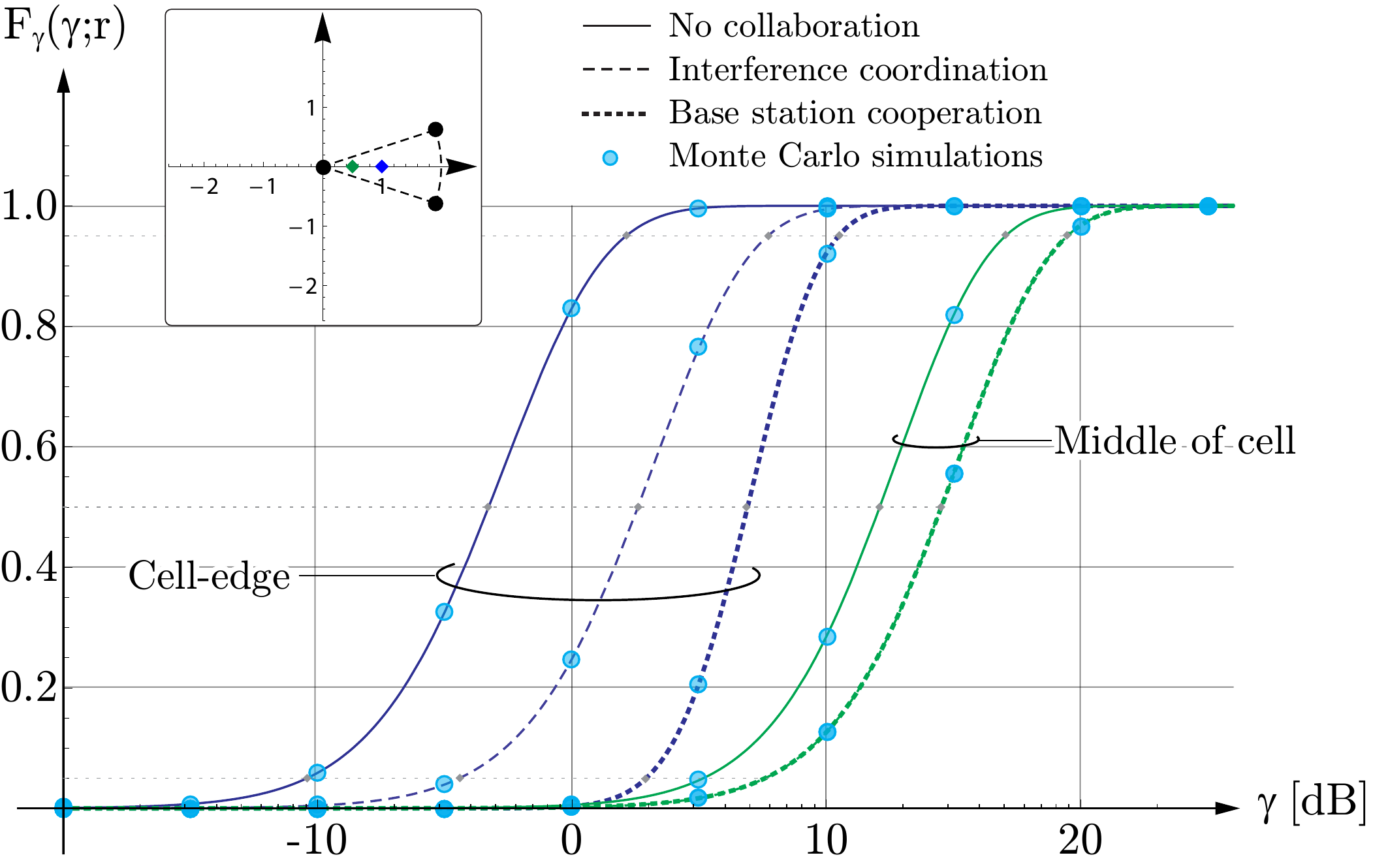}
	\caption{\ac{SIR} \ac{CDF} curves for user locations in the \emph{middle of the cell} ($r=0.5$) and at \emph{cell-edge} ($r=1$), respectively. Three cases are depicted: (i) No collaboration among \acp{BS} (solid), (ii) interference coordination (dashed), (iii) cooperation among \acp{BS} (dotted).}
	\label{Fig:SIR_Distributions}
\end{figure}

\begin{figure}
	\centering
	\includegraphics[width=.7\textwidth]{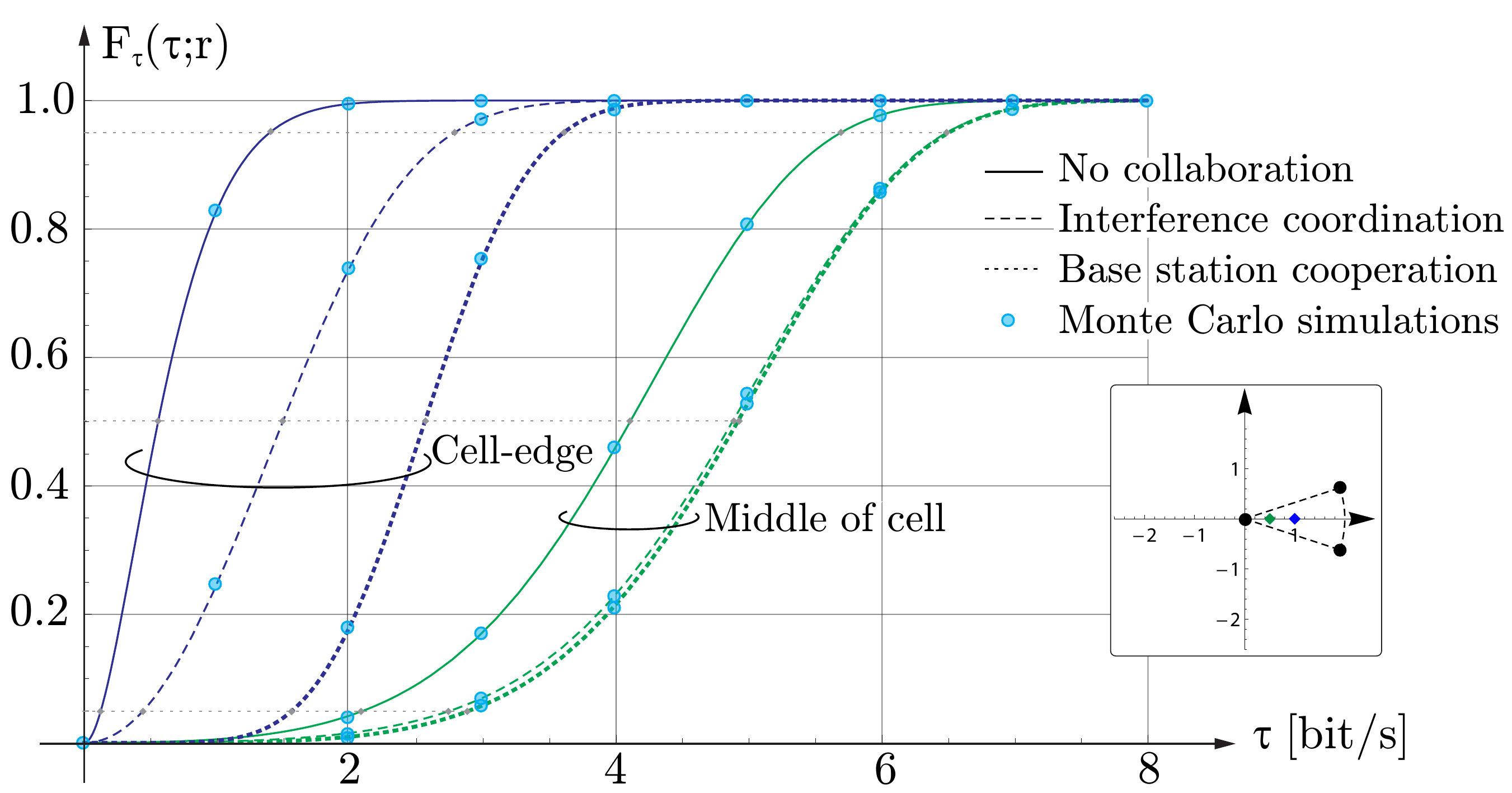}
	\caption{Normalized-rate \ac{CDF} curves for user locations in the \emph{middle of the cell} ($r=0.5$) and at \emph{cell-edge} ($r=1$), respectively. Three cases are depicted: (i) No collaboration among \acp{BS} (solid), (ii) interference coordination (dashed), (iii) cooperation among \acp{BS} (dotted).}
	\label{Fig:Rate_Distributions}
\end{figure}

The distributions $f_{\gamma}(\gamma;r)$ and $f_{\tau}(\tau;r)$ are analyzed at $r=0.5$ and $r=1$ referring to \emph{middle of the cell}, and \emph{cell-edge}, respectively. For reasons of clarity, \ac{CDF} curves are presented. In order to verify the analysis, Monte Carlo simulations are carried out, employing the system model from \cref{Sec:GenericCircularModel} and the signal propagation model from \cref{Tab:ParameterSetup}. The results are computed by averaging over $10^7$ channel realizations for each \ac{BS} collaboration scheme and each user location, and are denoted as bold dots in \cref{Fig:SIR_Distributions,Fig:Rate_Distributions}, respectively.

\cref{Fig:SIR_Distributions} shows the obtained \ac{SIR} distributions. It is observed that

\begin{itemize}
\item In the case of \emph{no collaboration} (solid lines in \cref{Fig:SIR_Distributions}), the curves have almost equal shape in the middle of the cell and at cell-edge. The distribution in the middle of the cell is slightly steeper due to the lower variance of the interferer impact. Their medians, hereafter used to represent the distributions' position, differ by 15.5\,dB. 

\item When the central node $\mT_{0,0}$ \textit{coordinates} its channel access with the user's two dominant interferers, $\mT_{1,1}$ and $\mT_{1,10}$, the \ac{SIR} improves by 2.4\,dB in the middle of the cell and 5.9\,dB at cell-edge (dashed curves in \cref{Fig:SIR_Distributions}), compared to \textit{no collaboration}.

\item \textit{\ac{BS} cooperation} enhances the \ac{SIR} by 10.2\,dB at cell-edge in comparison to \textit{no collaboration} (left dotted curve in \cref{Fig:SIR_Distributions}). Note that the \ac{CDF} curve also has a steeper slope than without coordination, indicating lower variance of the \ac{SIR}.

\item In the middle of the cell, \emph{cooperation} achieves hardly any additional improvement, as recognized from the overlapping rightmost curves in \cref{Fig:SIR_Distributions}. This remarkable result states that \emph{interference coordination} already performs close to optimal at this user location. Note that in realistic networks \emph{coordination} is typically far less complex than \emph{cooperation}.


\end{itemize}
The curves reflect findings from \citep{6482234}, stating that even in the best case, gains of transmitter cooperation are much smaller than largely envisioned. \cref{Fig:Rate_Distributions} depicts the corresponding rate distributions. The results show that
\begin{itemize}
\item Notably, the rate statistics of all three collaboration schemes indicate lower variance at cell-edge than in the middle of the cell.

\item In terms of median value, \emph{\ac{BS} coordination} shows rate improvements by $18.7\, \%$ in the middle of the cell and by $167\, \%$ at cell-edge.

\item \emph{Cooperation} between the central node $\mT_{0,0}$ and the user's two closest interferers, $\mT_{1,1}$ and $\mT_{1,10}$, achieves a rate enhancement of 19.8\,\% in the middle of the cell and 355.7\,\% at cell-edge. Similar to the \ac{SIR}, it is observed that in the middle of the cell, interference coordination already performs close to optimal.
\end{itemize}


In summary, collaboration among the \acp{BS} that were identified as main contributors to the shape of the interference distribution by \Cref{Thm:Theorem1}, achieved large performance enhancements in terms of \ac{SIR} and rate. 
It was further shown that the efficiency of such schemes considerably depends on the user eccentricity, or equivalently, the asymmetry of the interference impact. 

\section{Summary and Conclusions}
\label{Sec:SGN_Summary}


This paper introduced a new interference model that enables to represent substantially large interferer deployments by a well-defined circular structure in terms of interference statistics. The model applies angle-dependent power profiles, which require the specification of a mapping procedure. The presented scheme, despite not claimed to be optimal, achieved to accurately capture heterogeneous interferer deployments with hundreds or even thousands of base stations by a circular model with only several tens of nodes, reducing complexity substantially. Motivated by the desire to decompose the aggregate interference distribution into the contributions from the individual sources, a new representation for the sum of Gamma random variables with integer shape parameter was proposed. The approach enabled to identify candidate base stations for user-centric base station collaboration schemes and to predict the corresponding \ac{SIR}- and rate statistics at eccentric user locations. It was shown that the performance largely depends on the asymmetry of the interference impact. At the same time, power profiles of \ac{PPP} scenarios indicated the frequent presence of one or a few dominant interferers. Interference modeling based on stochastic geometry tends to conceal this diversity by spatial averaging and isotropic conditions. The authors are therefore confident that the introduced framework offers a convenient tool for investigating fifth generation mobile cellular networks, where the number of interferers and the diversity of scenarios is expected to grow substantially. Further work is mainly directed towards enhanced mapping schemes.

\appendix
\subsection{Proof of \Cref{Thm:Theorem1}}
\label{App:ProofOfTheorem}

Let $G_l \sim \Gamma[k_l, \theta_l]$ be $L$ independent random variables with $k_l$ being positive integers and all $\theta_l$ different. Then, the \ac{PDF} of $Y=\sum_{l=1}^L G_l$ can be expressed as \citep{6292935}
\begin{align}
f_Y(y) &= \left(\prod_{i=1}^{L} \frac{1}{\theta _i^{k_i}}\right)\frac{1}{2\pi \imath}\oint_C \frac{\prod_{i=1}^{L}\left\{\Gamma\left(\frac{1}{\theta _i}+s\right) \right\}^{k_i}}{\prod_{i=1}^{L}\left\{\Gamma \left(1+\frac{1}{\theta _i}+s\right)\right\}^{k_i}}e^{sy}ds  \label{eq:SumGammaMeijerG} \\
&=\left(\prod_{i=1}^{L} \frac{1}{\theta _i^{k_i}}\right) \mathcal{G}_{\mK,\mK}^{\mK,0}\left[e^{-y}\left|
\begin{array}{l}
\fT_a \\
\fT_b
\end{array}
\right.\right], \label{eq:MeijerGPDF}
\end{align}
where $\mathcal{G}_{p,q}^{m,n}[\cdot]$ denotes Meijer's G function, $\mK=\sum_{i=1}^L k_i$, and
\begin{align}
\fT_a&=\left\{\overbrace{\left(1+\frac{1}{\theta_1}\right),\ldots,\left(1+\frac{1}{\theta_1}\right)}^{k_1\   {\rm times}},\ldots ,\overbrace{\left(1+\frac{1}{\theta_L}\right),\ldots,\left(1+\frac{1}{\theta_L}\right)}^{k_L\   {\rm times}}\right\} \label{eq:aVector},\\
\fT_b&=\left\{\overbrace{\left(\frac{1}{\theta_1}\right),\ldots,\left(\frac{1}{\theta_1}\right)}^{k_1\   {\rm times}},\ldots,\overbrace{\left(\frac{1}{\theta_L}\right),\ldots,\left(\frac{1}{\theta_L}\right)}^{k_L\   {\rm times}}\right\} \label{eq:bVector}.
\end{align}
The unique values of $\fT_a$ and $\fT_b$ and their multiplicities $k_i$ are gathered by the vectors $\fa$, $\fb$ and $\fk$, respectively. Then, $|\fa|=|\fb|=L$, $a_i = \left(1+1/\theta_i\right)$ and $b_i=\left(1/\theta_i\right)$ for $i=1,\ldots, L$.

By virtue of the calculus of residues, \cref{eq:SumGammaMeijerG} can be evaluated by a summation over the negative residues of the integrand
\begin{equation}
I(s) = \frac{\prod_{i=1}^{L}\left\{\Gamma\left(\frac{1}{\theta_i}+s\right) \right\}^{k_i}}{\prod_{i=1}^{L}\left\{\Gamma \left(1+\frac{1}{\theta_i}+s\right)\right\}^{k_i}}z^s
\end{equation}
as
\begin{equation}
\mathcal{G}_{\mK,\mK}^{\mK,0}\left[z\left|
\begin{array}{l}
\fT_a \\
\fT_b
\end{array}
\right.\right] = -\sum_{l = 1}^{L} \sum_{j=0}^{\infty}R_l(j).
\label{eq:SumOfClasses}
\end{equation}

With
\begin{equation}
R_l(j) = \frac{1}{(k_l-1)!} \frac{d^{k_l-1}}{ds^{k_l -1}}\left.\left\{\left(s-\left(\frac{1}{\theta_l}+j\right)^{k_{l}}\right)I(s)\right\}\right|_{s=\frac{1}{\theta_l}+j}
\end{equation}
and the substitution $s = \frac{1}{\theta_l}+k+\zeta$, it is obtained
\begin{align}
R_l(j) &= \frac{1}{(k_l-1)!} \frac{d^{k_l-1}}{d\zeta^{k_l -1}} g(\zeta; j)=  g_l(0;j) \frac{h_{k_{l}-1}(0;j)}{(k_l-1)!}. \label{eq:residues}
\end{align}
Auxiliary function $g_l(0;j)$ is calculated as
\begin{align}
g_l(0;j) = & (-1)^{k_l} z^{1/\theta_l} \frac{\prod_{ i=1, i\neq l}^{L}\Gamma(\beta_i)^{k_i}}{\prod_{i = 1}^L \Gamma(\alpha_i)^{k_i}} \frac{z^j}{j!} \frac{\prod_{i = 1}^L\left((1-\alpha_i)_j\right)^{k_i}}{\prod_{i=1, i\neq l}^{L}\left((1-\beta_i)_j\right)^{k_i}},
\label{eq:helperg}
\end{align}
where $(\cdot)_c$ refers to the Pochhammer symbol, which is specified as $(x)_j = x(x+1)\ldots(x+j-1)$. The therms $\alpha_i$ and $\beta_i$ are defined as $\alpha_i = a_i - b_l$ and $\beta_i = b_i - b_l$, respectively. 

Auxiliary function $h_{\delta,l}(0;j)$ is recursively determined as
\begin{equation}
h_{\delta+1,l}(\zeta;j) = h_{1,l}(\zeta; j) h_{\delta,l}(\zeta;j)+\frac{d}{d\zeta}h_{\delta,l} (\zeta; j).
\label{eq:recHelper}
\end{equation}
It is left to provide the expressions for $h_{1,l}(\zeta; j)$ and $h_{1,l}^{(m)}(\zeta; j)$ at $\zeta=0$:
\begin{align}
h_{1,l}(0; j) =& \log(z) - k_l\, \psi(1 + j)  - \sum_{i=k_l+1}^\mK \psi(\beta_i-j) + \sum_{i=1}^\mK \psi(\alpha_i -j), \label{eq:h1} 
\end{align}
\begin{align}
h_{1,l}^{(m)}(0;j) =& \left.\frac{d^{m}}{d\zeta^m} h_{1,l}(\zeta; j) \right|_{\zeta=0}  \label{eq:h1tau} \nonumber\\
=& k_l \psi^{(m)}(1) - k_l \psi^{(m)}(1 + j) + (-1)^m\left(-k_l \psi^{(m)}(1) - \sum_{i=1, i\neq l}^{L} \psi^{(m)}(\beta_i-j) + \sum_{i=1}^L \psi^{(m)}(\alpha_i - j)\right), 
\end{align}
where $\psi^{(m)}(z) = \frac{d^m}{dz^m} \log\left(\Gamma(z)\right)$ refers to the polygamma function of order $m$.

Since $\alpha_i = 1$ for $i=l$, the argument $(\alpha_i - j)$ in \cref{eq:h1,eq:h1tau} can take on non-positive integer values for $j>1$, where the polygamma function has poles of order $m+1$.
These poles are however compensated by the zeros $(1-\alpha_i)$ in \cref{eq:helperg} due to the following facts: (i) By definition, $(0)_c=0$ for $j\geq1$, (ii) the zeros are of order $k_l$ and (iii) for any non-positive integer $q$
\begin{align}
\lim\limits_{x\rightarrow q}\,  (x-q)^{k_l} \psi^{(k_l-2)}(x-q) = 0 \label{eq:limDerivation},\\
\lim\limits_{x\rightarrow q}\,  (x-q)^{k_l} \left(\psi^{(0)}(x-q)\right)^{k_l-1} = 0.
\label{eq:limExp}
\end{align}
The derivation order $(k_l - 2)$ and the exponent $k_l-1$ in \cref{eq:limDerivation,eq:limExp} correspond to the respective maximum values in $h_{k_l-1,l}(0;j)$.
Consequently, $R_l(j)=0 $ for $ j>0$ and, therefore, \cref{eq:SumOfClasses} is simplified as
\begin{equation}
\mathcal{G}_{\mK,\mK}^{\mK,0}\left[z\left|
\begin{array}{l}
\fT_a \\
\fT_b
\end{array}
\right.\right] = -\sum_{l = 1}^{L} R_l(0).
\end{equation}
$R_l(0)$ is composed of $h_{\delta,l}(0;0)$ and $g_l (0;0)$. 

From \cref{eq:h1,eq:h1tau} it holds that
\begin{align}
h_{1,l}(0;0) &= \log(z) + \sum_{i=1, i\neq l}^{L}\frac{1}{\beta_i} \label{eq:h1simp},\\
h_{1,l}^{(m)}(0;0) &=  m! \sum_{i=1, i\neq l}^{L}\left(\frac{1}{\beta_i}\right)^{m+1}\label{eq:h1tausimp},
\end{align}
where the recurrence relation of the Polygamma function is applied. Simple manipulations yield \cref{eq:h1t,eq:h1taut}. With \cref{eq:htau}, $h_{\delta,l}(0;0)$ can be derived. Note that in \cref{eq:h1t,eq:h1taut,eq:htau} the second "$0$" in the argument, which stems from $j=0$, is omitted for readability.

Considering that $\alpha_i = 1 + \beta_i$ and using the recurrence relation $\Gamma(z+1) = z\Gamma(z)$ of the Gamma function, \cref{eq:helperg} can be simplified as
\begin{equation}
g(0;0) = (-1)^{k_l} z^{1/\theta_l} \prod_{i=1, i\neq l}^{L} \left(\frac{1}{\beta_i}\right)^{k_i}.
\label{Eq:helpergsimplified}
\end{equation}

Finally, \cref{Eq:LambdaEll,Eq:Theorem1} are obtained from \cref{eq:h1simp,eq:h1tausimp,Eq:helpergsimplified}.


\small
\bibliographystyle{alpha}
\bibliography{SumOfGammaRandomVariables}

\newcommand{\etalchar}[1]{$^{#1}$}
\begin{thebibliography}{MWMZ07}

\bibitem[3GP13a]{3gpp.36.819}
3GPP.
\newblock Coordinated multi-point operation for {LTE} physical layer aspects.
\newblock TR {36.819}, {3rd Generation Partnership Project (3GPP)}, Sept. 2013.

\bibitem[3GP13b]{3gpp.36.839}
3GPP.
\newblock {Evolved Universal Terrestrial Radio Access} ({E-UTRA}); mobility
  enhancements in heterogeneous networks.
\newblock TR {36.839}, {3rd Generation Partnership Project (3GPP)}, Jan. 2013.

\bibitem[AAK01]{966578}
M.-S. Alouini, A.~Abdi, and Mostafa Kaveh.
\newblock Sum of {G}amma variates and performance of wireless communication
  systems over {N}akagami-fading channels.
\newblock {\em IEEE Transactions on Vehicular Technology}, 50(6):1471--1480,
  Nov. 2001.

\bibitem[ACD{\etalchar{+}}12]{6171992}
J.G. Andrews, H.~Claussen, M.~Dohler, S.~Rangan, and M.C. Reed.
\newblock Femtocells: Past, present, and future.
\newblock {\em IEEE Journal on Selected Areas in Communications},
  30(3):497--508, April 2012.

\bibitem[ADB94]{abu1994outage}
Adnan~A Abu-Dayya and Norman~C Beaulieu.
\newblock Outage probabilities in the presence of correlated lognormal
  interferers.
\newblock {\em IEEE Transactions on Vehicular Technology}, 43(1):164--173, Feb.
  1994.

\bibitem[AGH{\etalchar{+}}10]{5621983}
J.G. Andrews, R.K. Ganti, M.~Haenggi, N.~Jindal, and S.~Weber.
\newblock A primer on spatial modeling and analysis in wireless networks.
\newblock {\em IEEE Communications Magazine}, 48(11):156--163, Nov. 2010.

\bibitem[AHAB85]{1096243}
Emad~K. Al-Hussaini and A.~Al-Bassiouni.
\newblock Performance of {MRC} diversity systems for the detection of signals
  with {N}akagami fading.
\newblock {\em IEEE Transactions on Communications}, 33(12):1315--1319, Dec.
  1985.

\bibitem[AM97]{693785}
S.V. Amari and R.B. Misra.
\newblock Closed-form expressions for distribution of sum of exponential random
  variables.
\newblock {\em IEEE Transactions on Reliability}, 46(4):519--522, Dec. 1997.

\bibitem[APE05]{1556824}
V.A. Aalo, T.~Piboongungon, and G.P. Efthymoglou.
\newblock Another look at the performance of {MRC} schemes in {N}akagami-m
  fading channels with arbitrary parameters.
\newblock {\em IEEE Transactions on Communications}, 53(12):2002--2005, Dec.
  2005.

\bibitem[AYAK12]{6292935}
I.S. Ansari, F.~Yilmaz, M.-S. Alouini, and O.~Kucur.
\newblock On the sum of {G}amma random variates with application to the
  performance of {Maximal Ratio Combining} over {N}akagami-m fading channels.
\newblock In {\em IEEE International Workshop on Signal Processing Advances in
  Wireless Communications (SPAWC)}, pages 394--398, June 2012.

\bibitem[BADM95]{beaulieu1995estimating}
Norman~C Beaulieu, Adnan~A Abu-Dayya, and Peter~J McLane.
\newblock Estimating the distribution of a sum of independent lognormal random
  variables.
\newblock {\em IEEE Transactions on Communications}, 43(12):2869--2873, Dec.
  1995.

\bibitem[BB09a]{BaccelliVolI}
Francois Baccelli and Bartllomiej Blaszczyszyn.
\newblock {\em Stochastic Geometry and Wireless Networks: Volume {I} Theory}.
\newblock Foundation and Trends in Networking. Now Publishers, March 2009.

\bibitem[BB09b]{BaccelliVolII}
Fran{\c c}ois Baccelli and Bartlomiej Blaszczyszyn.
\newblock {\em Stochastic Geometry and Wireless Networks, Volume {II} -
  Applications}, volume~2 of {\em {Foundations and Trends in Networking}}.
\newblock NoW Publishers, 2009.

\bibitem[BKLZ97]{StochasticGeomAndArchi}
François Baccelli, Maurice Klein, Marc Lebourges, and Sergei Zuyev.
\newblock Stochastic geometry and architecture of communication networks.
\newblock {\em Telecommunication Systems}, 7(1-3):209--227, 1997.

\bibitem[BLM{\etalchar{+}}14]{6736747}
N.~Bhushan, Junyi Li, D.~Malladi, R.~Gilmore, D.~Brenner, A.~Damnjanovic,
  R.~Sukhavasi, C.~Patel, and S.~Geirhofer.
\newblock Network densification: the dominant theme for wireless evolution into
  {5G}.
\newblock {\em IEEE Communications Magazine}, 52(2):82--89, Feb. 2014.

\bibitem[Bro00]{895048}
T.X. Brown.
\newblock Cellular performance bounds via shotgun cellular systems.
\newblock {\em IEEE Journal on Selected Areas in Communications},
  18(11):2443--2455, Nov. 2000.

\bibitem[BVH14]{BaiBlockage}
T.~Bai, R.~Vaze, and R.~{Heath, Jr.}
\newblock Analysis of blockage effects on urban cellular networks.
\newblock {\em IEEE Transactions on Wireless Communications}, 13(9):5070--5083,
  Sept. 2014.

\bibitem[BZ96]{Baccelli96stochasticgeometry}
François Baccelli and Serguei Zuyev.
\newblock Stochastic geometry models of mobile communication networks.
\newblock In {\em Frontiers in queueing: models and applications in science and
  engineering}, pages 227--243. CRC Press, 1996.

\bibitem[Coe98]{Coelho199886}
Carlos~A. Coelho.
\newblock The generalized integer {G}amma distribution - a basis for
  distributions in multivariate statistics.
\newblock {\em Journal of Multivariate Analysis}, 64(1):86 -- 102, 1998.

\bibitem[Cur41]{1941}
J.~H. Curtiss.
\newblock On the distribution of the quotient of two chance variables.
\newblock {\em The Annals of Mathematical Statistics}, 12(4):409--421, 1941.

\bibitem[DRGC13]{6516171}
M.~Di~Renzo, A~Guidotti, and G.E. Corazza.
\newblock Average rate of downlink heterogeneous cellular networks over
  generalized fading channels: A stochastic geometry approach.
\newblock {\em IEEE Transactions on Communications}, 61(7):3050--3071, July
  2013.

\bibitem[EHH13]{elsawy2013stochastic}
Hesham ElSawy, Ekram Hossain, and Martin Haenggi.
\newblock Stochastic geometry for modeling, analysis, and design of multi-tier
  and cognitive cellular wireless networks: A survey.
\newblock {\em IEEE Communications Surveys \& Tutorials}, 15(3):996--1019, June
  2013.

\bibitem[EPA06]{1583918}
G.P. Efthymoglou, T.~Piboongungon, and V.A. Aalo.
\newblock Performance of {DS-CDMA} receivers with {MRC} in {N}akagami-m fading
  channels with arbitrary fading parameters.
\newblock {\em IEEE Transactions on Vehicular Technology}, 55(1):104--114, Jan.
  2006.

\bibitem[Ger12]{BellLabsIdeaFactory}
Jon Gertner.
\newblock {\em The Idea Factory: {B}ell {L}abs and the Great Age of American
  Innovation}.
\newblock Penguin Group, 2012.

\bibitem[GH13]{guo2013spatial}
Anjin Guo and Martin Haenggi.
\newblock Spatial stochastic models and metrics for the structure of base
  stations in cellular networks.
\newblock {\em IEEE Transactions on Wireless Communications},
  12(11):5800--5812, Oct. 2013.

\bibitem[GMR{\etalchar{+}}12]{ghosh2012heterogeneous}
Amitabha Ghosh, Nitin Mangalvedhe, Rapeepat Ratasuk, Bishwarup Mondal, Mark
  Cudak, Eugene Visotsky, Timothy~A Thomas, Jeffrey~G Andrews, Ping Xia,
  Han~Shin Jo, et~al.
\newblock Heterogeneous cellular networks: From theory to practice.
\newblock {\em IEEE Communications Magazine}, 50(6):54--64, June 2012.

\bibitem[HAB{\etalchar{+}}09]{5226957}
M.~Haenggi, J.G. Andrews, F.~Baccelli, O.~Dousse, and M.~Franceschetti.
\newblock Stochastic geometry and random graphs for the analysis and design of
  wireless networks.
\newblock {\em IEEE Journal on Selected Areas in Communications},
  27(7):1029--1046, Sept. 2009.

\bibitem[Hae12]{haenggi2012stochastic}
M.~Haenggi.
\newblock {\em Stochastic Geometry for Wireless Networks}.
\newblock Stochastic Geometry for Wireless Networks. Cambridge University
  Press, 2012.

\bibitem[HB05]{hu2005accurate}
Jeremiah Hu and Norman~C Beaulieu.
\newblock Accurate simple closed-form approximations to {R}ayleigh sum
  distributions and densities.
\newblock {\em IEEE Communications Letters}, 9(2):109--111, Feb. 2005.

\bibitem[HG09]{Haenggi:2009}
Martin Haenggi and Radha~Krishna Ganti.
\newblock {\em Interference in Large Wireless Networks}, volume~3 of {\em
  Foundations and Trends in Networking}.
\newblock NoW Publishers, Feb. 2009.

\bibitem[HKB13]{6515339}
Robert~W. {Heath, Jr.}, Marios Kountouris, and Tianyang Bai.
\newblock Modeling heterogeneous network interference using poisson point
  processes.
\newblock {\em IEEE Transactions on Signal Processing}, 61(16):4114--4126, Aug.
  2013.

\bibitem[IM13]{mabrouk}
Serge B.~Provost Iman~Mabrouk.
\newblock The exact density function of a sum of independent {G}amma random
  variables as an inverse {M}ellin transform.
\newblock {\em International Journal of Applied Mathematics and Statistics},
  41(11), 2013.

\bibitem[Kab62]{1962}
D.~G. Kabe.
\newblock On the exact distribution of a class of multivariate test criteria.
\newblock {\em The Annals of Mathematical Statistics}, 33(3):1197--1200, 1962.

\bibitem[KST06]{1673666}
G.K. Karagiannidis, N.C. Sagias, and T.A. Tsiftsis.
\newblock Closed-form statistics for the sum of squared {N}akagami-m variates
  and its applications.
\newblock {\em IEEE Transactions on Communications}, 54(8):1353--1359, Aug.
  2006.

\bibitem[LFR99]{747812}
P.~Lombardo, G.~Fedele, and M.M. Rao.
\newblock {MRC} performance for binary signals in {N}akagami fading with
  general branch correlation.
\newblock {\em IEEE Transactions on Communications}, 47(1):44--52, Jan. 1999.

\bibitem[LHA13]{6482234}
A.~Lozano, R.W. Heath, and J.G. Andrews.
\newblock Fundamental limits of cooperation.
\newblock {\em IEEE Transactions on Information Theory}, 59(9):5213--5226,
  Sept. 2013.

\bibitem[Mol12]{moltchanov2012distance}
Dmitri Moltchanov.
\newblock Distance distributions in random networks.
\newblock {\em Ad Hoc Networks}, 10(6):1146--1166, March 2012.

\bibitem[Mos85]{moschopoulos}
P.G. Moschopoulos.
\newblock The distribution of the sum of independent {G}amma random variables.
\newblock {\em Annals of the Institute of Statistical Mathematics},
  37(1):541--544, 1985.

\bibitem[MWMZ07]{mehta2007approximating}
Neelesh~B Mehta, Jingxian Wu, Andreas~F Molisch, and Jin Zhang.
\newblock Approximating a sum of random variables with a lognormal.
\newblock {\em IEEE Transactions on Wireless Communications}, 6(7):2690--2699,
  July 2007.

\bibitem[PDL06]{PDL06:SCVT}
Simon Plass, Xenofon~G. Doukopoulos, and Rodolphe Legouable.
\newblock Investigations on link-level inter-cell interference in {OFDMA}
  systems.
\newblock In {\em Symposium on Communications and Vehicular Technology (SCVT)},
  pages 49--52, Nov. 2006.

\bibitem[PK91]{289422}
R.~Prasad and A~Kegel.
\newblock Improved assessment of interference limits in cellular radio
  performance.
\newblock {\em IEEE Transactions on Vehicular Technology}, 40(2):412--419, May
  1991.

\bibitem[Pro89]{provost}
Serge~B. Provost.
\newblock On sums of independent {G}amma random variates.
\newblock {\em Statistics: A Journal of Theoretical and Applied Statistics},
  20(4), 1989.

\bibitem[RZXZ13]{6747823}
Chengmeng Ren, Jianfeng Zhang, Wei Xie, and Dongmei Zhang.
\newblock Performance analysis for heterogeneous cellular networks based on
  {M}atern-like point process model.
\newblock In {\em International Conference on Information Science and
  Technology (ICIST)}, pages 1507--1511, March 2013.

\bibitem[Sch88]{3717}
E.M. Scheuer.
\newblock Reliability of an m-out-of-n system when component failure induces
  higher failure rates in survivors.
\newblock {\em IEEE Transactions on Reliability}, 37(1):73--74, April 1988.

\bibitem[TDHA14]{6779696}
H.~Tabassum, Z.~Dawy, E.~Hossain, and M.-S. Alouini.
\newblock Interference statistics and capacity analysis for uplink transmission
  in two-tier small cell networks: A geometric probability approach.
\newblock {\em IEEE Transactions on Wireless Communications}, 13(7):3837--3852,
  July 2014.

\bibitem[TKKS06]{C:Sagias_C2_2006}
T.~A. Tsiftsis, G.~K. Karagiannidis, S.~A. Kotsopoulos, and N.~C. Sagias.
\newblock Performance of {MRC} diversity receivers over correlated
  {N}akagami-$m$ fading channels.
\newblock In {\em {I}nternational {S}ymposium of {C}ommunication {S}ystems,
  {N}etworks and {D}igital {S}ignal {P}rocessing ({CSNDSP})}, July 2006.

\bibitem[TV12]{TorrieriV12}
Don~J. Torrieri and Matthew~C. Valenti.
\newblock The outage probability of a finite ad hoc network in {N}akagami
  fading.
\newblock {\em IEEE Transactions on Communications}, 60(11):3509--3518, Nov.
  2012.

\bibitem[TYDA13]{6378490}
H.~Tabassum, F.~Yilmaz, Z.~Dawy, and M.-S. Alouini.
\newblock A framework for uplink intercell interference modeling with
  channel-based scheduling.
\newblock {\em IEEE Transactions on Wireless Communications}, 12(1):206--217,
  Jan. 2013.

\bibitem[WA12]{NET-032}
Steven Weber and Jeffrey~G. Andrews.
\newblock Transmission capacity of wireless networks.
\newblock {\em Foundations and Trends in Networking}, 5(2-3):109--281, 2012.

\bibitem[WPS09]{4802198}
M.Z. Win, P.C. Pinto, and L.A Shepp.
\newblock A mathematical theory of network interference and its applications.
\newblock {\em Proceedings of the IEEE}, 97(2):205--230, Feb. 2009.

\bibitem[Wyn94]{340450}
AD. Wyner.
\newblock Shannon-theoretic approach to a {G}aussian cellular multiple-access
  channel.
\newblock {\em IEEE Transactions on Information Theory}, 40(6):1713--1727, Nov.
  1994.

\bibitem[YC08]{4524273}
Lie-Liang Yang and Hsiao-Hwa Chen.
\newblock Error probability of digital communications using relay diversity
  over {N}akagami-m fading channels.
\newblock {\em IEEE Transactions on Wireless Communications}, 7(5):1806--1811,
  May 2008.

\bibitem[Zha98]{729390}
Q.T. Zhang.
\newblock Exact analysis of postdetection combining for {DPSK} and {NFSK}
  systems over arbitrarily correlated {N}akagami channels.
\newblock {\em IEEE Transactions on Communications}, 46(11):1459--1467, Nov.
  1998.

\end{thebibliography}

\end{document}